\documentstyle[12pt,epsfig,bbm,amsmath,amssymb,verbatim]{article}

\begin{document}

\newcommand{\id}{{\hbox{{\rm 1}\kern-.26em\hbox{\rm l}}}}
\newcommand{\CC}{{\mathbb {C}}}
\newcommand{\ZZ}{{\mathbb {Z}}}
\newcommand{\ord}{\mbox{{\rm ord}\,}}

\def\topfraction{0.9}
\def\bottomfraction{0.9}
\def\intextfraction{0.9}
\def\textfraction{0.1}

\def\printfidsym#1{\penalty-100\noindent\textbf{\boldmath$#1$}\penalty10000\begin{scriptsize}\verbatiminput{sicsymbolic_#1.txt}\end{scriptsize}\vskip20pt plus 10pt minus 10pt}

\begin{center}

{\Large The monomial representations of the Clifford group}

\vspace{12mm}

D.M.\ Appleby$^1$
\hspace{10mm} 
Ingemar Bengtsson$^2$
\hspace{10mm} 
Stephen Brierley$^{3,\, 4}$

\

\

Markus Grassl$^5$ 
\hspace{10mm} 
David Gross$^6$ 
\hspace{10mm} 
Jan-\AA ke Larsson$^7$ 

\vspace{14mm}

{\small 
$^1${\it Perimeter Institute for Theoretical Physics, 31 Caroline 
Street North, Waterloo, Ontario N2L 2Y5, Canada}

$^2${\it Stockholms Universitet, AlbaNova, Fysikum, 
S-106 91 Stockholm, Sweden}

$^3${\it Department of Mathematics, University of Bristol, Bristol BS8 1TW, UK}

$^4${\it QuIC, Ecole Polytechnique, Universit\'e Libre de Bruxelles, CP 165, 1050 Brussels, Belgium}

$^5${\it Centre for Quantum Technologies, National University of Singapore,
Singapore~117543} 

$^6${\it Institute for Theoretical Physics, ETH Z\"urich, 8093 
Z\"urich, Switzerland}

$^7${\it Institutionen f\"or Systemteknik och Matematiska Institutionen, 
Link\"opings Universitet, S-581 83 Link\"oping, Sweden.}
}
\vspace{10mm}

{\bf Abstract:}

\end{center}

\

\noindent We show that the Clifford group---the normaliser of the 
Weyl-Heisenberg group---can be represented by monomial phase-permutation 
matrices if and only if the dimension is a square number. This simplifies 
expressions for SIC vectors, and has other applications to SICs and to Mutually Unbiased Bases. Exact solutions for SICs in dimension 16 are presented for the first time.

\vspace{15mm}

\begin{center}

%

\end{center}

\newpage

{\bf 1. Introduction}

\vspace{5mm}

\noindent The Weyl-Heisenberg group \cite{Weyl} first appeared in 
nineteenth century algebraic geometry, and is at the roots of many things including harmonic analysis, theta functions, and---of course---quantum 
mechanics. Its automorphism group within the unitary group---the 
largest subgroup of the unitary group having the Weyl-Heisenberg group 
as a normal subgroup---appears in quantum information theory under 
the name of the Clifford group \cite{Fivel, Gottesman}. 

The particular problem that motivated the present study 
is known as the SIC problem \cite{Zauner, Renes}: in a complex Hilbert space of finite dimension $N$, 
find $N^2$ unit vectors $|\psi_I\rangle $ such that 

\begin{equation} I \neq J \hspace{5mm} \Rightarrow \hspace{5mm} 
|\langle \psi_I|\psi_J \rangle |^2 = \frac{1}{N+1} \ . 
\end{equation}

\noindent Such a collection of equiangular vectors \cite{equi} is
known as a SIC, which is short for a Symmetric Informationally
Complete POVM, where POVM is short for Positive Operator Valued
Measure. In physics a SIC represents a kind of fiducial measurement,
of interest for quantum state tomography \cite{Renes} and indeed for
the very foundations of quantum mechanics \cite{RS,kub,Asa}. Such
measurements can be---and in low dimensions, have been \cite{Durt,
  Durt2, Steinberg}---realised in the laboratory, but---perhaps
surprisingly---the theoretical SIC problem as stated is not easy to
solve. At the moment exact solutions are known in dimensions 2--16,
19, 24, 35, and 48 (with 16 added here), while convincing numerical
solutions are available in dimensions 2--67. These results are due to
several authors; we refer to Scott and Grassl \cite{Scott} for
complete references, and just remark that beyond three dimensions the
known SICs look appallingly complicated at first sight.

Zauner's conjecture \cite{Zauner} states that in every dimension 
there is an orbit of the Weyl-Heisenberg group which forms a SIC, and 
moreover that every vector in such a SIC is left invariant by an element 
of the Clifford group of order three. Hence the problem of finding 
a SIC reduces to that of finding a suitable fiducial vector for the group 
to act on, and the second---very mysterious---part of the conjecture provides 
some guidance when one looks for such a fiducial vector. All available 
evidence supports Zauner's conjecture \cite{Scott, Marcus}.

The representation theory of the Weyl-Heisenberg group tells us that 
once one of its generators is given in diagonal form all its 
group elements are represented by monomial phase-permutation 
matrices, that is by unitary matrices having only one non-zero 
entry per column, and per row \cite{Weyl}. Being products of 
a permutation matrix and a diagonal unitary, such matrices are 
also said to be of shift-and-multiply type. This property of 
the group can be traced back to the way that the representation 
is induced from that of its center. It is an 
important property shared by many, but not all, unitary operator 
bases of group type (or ``nice error bases'', as they are called 
in quantum information theory) \cite{Klappenecker}. 

In general the Clifford group is not represented by phase-permutation 
matrices. However, after a few preliminaries in section 2, 
we devote section 3 to a representation of the Weyl-Heisenberg 
group which is special to the case when the dimension is 
a square, $N = n^2$. It can be thought of as a finite dimensional 
analogue of the Zak basis \cite{Zak}, and is used in the theory of 
theta functions \cite{Mumford}. Our observation is that in this representation the entire Clifford group 
is given by phase-permutation matrices. In section 4 we demonstrate 
that this remarkable feature is present if and only if the dimension is 
a square. In the remaining sections we explore 
some ways in which the phase-permutation basis can be useful. Section 
5 illustrates how it is, in a way, aligned to SICs, while section 6 
is devoted to exact solutions for SICs in $2^2$ and $3^2$ dimensions. 
In the former case they are trivial to obtain. In the latter 
they are not, but they look significantly better compared 
to how they look in the standard basis \cite{Scott, Grassl}. 
In Section 7, we present new exact solutions for dimesion $4^2$; previously these were known in numerical form only. Section 8 contains a remark on Mutually Unbiased Bases, and section 9 summarises our conclusions. There are three appendices containing group theoretical theorems and the fiducial vectors for dimension $4^2$.

\vspace{10mm}

{\bf 2. Preliminaries}

\vspace{5mm}

\noindent We introduce the \emph{Weyl-Heisenberg group} by writing
down a defining representation. Choose a dimension $N$ and assume that
$\{|0\rangle, \dots, |N-1\rangle\}$ is an orthonormal basis of
$\CC^N$. Define two phase factors $\omega, \tau$ by
\begin{equation} 
	\omega = e^{\frac{2\pi i}{N}} \ , \hspace{8mm} 
	\tau = - e^{\frac{i\pi }{N}} 
\end{equation}
and two operators $X, Z$ by
\begin{equation} 
	X|u\rangle = |u+1 \rangle \ , \hspace{8mm} 
	Z|u\rangle = \omega^u|u\rangle \ , \hspace{5mm} u \in \{ 0, \dots , N-1\} \ , 
\label{Steve's7} \end{equation}
where here and elsewhere the labels of the vectors are computed modulo $N$. The matrix group generated by $\{\tau, X, Z\}$ is the defining
representation of the Weyl-Heisenberg group $H(N)$. (It is known
\cite{Weyl} that all irreducible representations of $H(N)$ in
dimensions larger than one are unitarily equivalent to the defining one).
Note that $X$ and $Z$ are represented by phase-permutation matrices,
and the same is true for all the elements of the
group since the product of two phase-permutation matrices is again a
phase-permutation matrix.  

The Weyl-Heisenberg groups $H(N)$ behave slightly differently
depending on whether $N$ is even or odd.  The underlying reason is
that while one always has the relations
\begin{equation} \label{eqn:orderN}
	\omega^N = 1\ , \qquad
	X^N = Z^N = {\bf 1}  \ ,
\end{equation}
the order of $\tau$ depends on the parity of $N$: 
\begin{equation}
	\tau^N =
	\left\{
		\begin{array}{ll}
			-1 \quad & N \ \mbox{even} \\
			1 \quad & N \ \mbox{odd}.
		\end{array}
	\right.
\end{equation}
As a consequence we will sometimes end up using arithmetic modulo $N$
in the odd case and modulo $2 N$ in the even case.
To unify the notation, we introduce the symbol
\begin{equation} 
	\bar{N} = \left\{ \begin{array}{ll} N & N \ 
\mbox{is odd} \\ 
2N & N \ \mbox{is even} \end{array} \right. \ . 
\end{equation}

\noindent Since $ZX = \omega XZ$, the introduction of the phase factor $\tau$ may 
seem odd. If $N$ is odd, $\tau$ is in fact a power of $\omega$. The reason it is 
included in the even dimensional case can be traced back to 
the fact that there are group elements, such as $XZ$, that generate cyclic subgrops of order $2N$. For us it 
will be crucial that the Clifford group defined below acts on $H(N)$ as 
we have defined it here. 

To analyze the structure of the Weyl-Heisenberg group, we define the
group elements
\begin{equation} \label{eqn:Dij}
	D_{ij} = \tau^{ij}X^iZ^j \ . 
\end{equation}  
for $i,j = 0, \dots, \bar N - 1$. 
One can then verify the central
\emph{composition law}
\begin{equation} 
	D_{ij}D_{lm} = \tau^{lj-im}D_{i+l,j+m}.   \label{symp} 
\end{equation}
Because the (non-scalar) generators $X, Z$ of $H(N)$ are just $D_{1
0}$ and $D_{0 1}$ respectively, the composition law (\ref{symp})
implies that any element of $H(N)$ is of the form $\tau^k
D_{ij}$ for suitable integers $i,j,k$. 

Note that the phase factor $\tau^{ij}$ in (\ref{eqn:Dij}) depends on
$ij$ modulo $\bar N$, whereas $X^i Z^j$ only depends on $i$ and $j$
modulo $N$ (by virtue of (\ref{eqn:orderN})). Hence the group law
(\ref{symp}) says that $H(N)$ modulo phase factors is isomorphic
to $\ZZ_N^2$, where $\ZZ_N$ is the group of integers $\{0, \dots,
N-1\}$ with addition modulo $N$.

The situation encountered in the last paragraph will become a general
theme below. Depending on our objective, we will take one of two
points of view. Sometimes we will be concerned with concrete matrix
representations of the groups involved. Then, we will be specific
about all phase factors involved, and work with arithmetic modulo
$\bar N$. In other situations, however, a more abstract approach turns
out to be beneficial.  In these cases, we will factor out phases and
work solely in terms of the discrete group $\ZZ_N^2$ and its symmetry
groups.

To make the abstract approach more precise, let $Z(N)$ be
the center of the Heisenberg-Weyl group $H(N)$. 
From the group law (\ref{symp}), it is evident that $Z(N) = \{
\tau^k\,\id \}_{k=0, \dots, \bar N-1}$, its elements are precisely the phase factors
times the identity matrix. Our previous observation can now be phrased
more succinctly as
\begin{equation}\label{eqn:factorGroup}
	H(N)/Z(N) \simeq \ZZ_N^2.
\end{equation}


\noindent Unless $N$ is a prime, the integers modulo $N$ form a ring but
not a field. Therefore, strictly speaking, the ``vectors''
$(i,j)\in\ZZ_{N}^2$ are elements of a module rather than of a vector
space. But we permit ourselves a slight abuse of terminology and speak
of vectors in $\ZZ_N^2$.

We will be concerned with another group: the \emph{Clifford group}.
It consists of all unitary operators $U_G$ normalising the
Weyl-Heisenberg group, in the sense that for all 
$i,j$ 
there are $i',j',k'$ such that
\begin{equation} 
	U_GD_{ij}U^{\dagger}_G = \tau^{k'} D_{i',j'} \ . 
	\label{defcliff} 
\end{equation}
Not all transformations $(i,j)\mapsto(i',j')$ are possible.
The fact that the group law (\ref{symp}) involves addition of vectors
in $\ZZ_{\bar N}^2$ suggests that any such transformation must be
linear. Further, the fact that the symplectic inner product $(lj-im)$
modulo $\bar N$ of the vectors $(i,j)$ and $(l, m)$ appears in
(\ref{symp}) suggests that this inner product might be an invariant
\begin{equation}\label{eqn:invariant}
	l j - i m = l'j' - i'm' \quad \mbox{mod}\ \bar N
\end{equation}
of the action of the Clifford group.

These intuitions turn out to be true and yield an almost exhaustive
understanding of the Clifford group \cite{Marcus}.  More precisely,
recall that $SL(2,\bar N)$ is the
group of linear transformations on $\ZZ_{\bar N}$ leaving the
symplectic inner product invariant. A $\ZZ_{\bar N}$-valued matrix 
\begin{equation}\label{eqn:Gmatrix}
	G = \left( \begin{array}{cc} \alpha & \beta \\ 
	\gamma & \delta \end{array} \right)
	\end{equation}	
is an element of $SL(2,\bar N)$ if and only if 
\begin{equation}
	\det G = 
	\alpha \delta - \beta \gamma = 1 \quad \mbox{mod} \ \bar{N} \ . 
\end{equation}
In one direction, we have that for every $G\in SL(2,\bar N)$, there
is an element $U_G$ in the Clifford group such that
\begin{equation}\label{eqn:UG}
		U_G D_{i j} U_G^\dagger = D_{G \left( i \atop j \right)}.
\end{equation}
A converse statement will be given below. The unitaries appearing in
(\ref{eqn:UG})  are known explicitly (c.f.\ Ref.~\cite{Marcus} for
more details): if $\beta$ is relatively prime to $\bar{N}$ ---so that
it has a multiplicative inverse---one finds  

\begin{equation} U_G = \frac{1}{\sqrt{N}}e^{i\theta}
\sum_{u,v=0}^{N-1}\tau^{\beta^{-1}(\delta u^2 - 2uv 
+\alpha v^2)}|u\rangle \langle v| \ , \label{standardrep} 
\end{equation}
where $e^{i\theta}$ is an arbitrary phase. If $\beta$ is not relatively prime to $\bar{N}$ we 
can use the decomposition  
\begin{equation} \left( \begin{array}{cc} \alpha & \beta \\ 
\gamma & \delta \end{array} \right) 
= \left( \begin{array}{cc} 0 & -1 \\ 1 & x 
\end{array}\right) \left( \begin{array}{cc} \gamma + x\alpha & 
\delta + x\beta \\ - \alpha & - \beta \end{array} \right) \ , \label{decomp} \end{equation}
where the integer $x$ can always be chosen so that 
$\delta + x\beta$ is relatively prime to $\bar{N}$. 
We remark that in this representation symplectic matrices are
represented by unitary phase-permutation matrices if and only if they
are of the form

\begin{equation} G = \left( \begin{array}{cc} \alpha & 0 \\ 
\gamma & \alpha^{-1} \end{array} \right)\ . \label{ZaunerG} 
\end{equation}

\noindent They form a rather small subgroup.

Let us return to the more abstract point of view alluded to before.
Since phase factors are left invariant by a unitary conjugation
\begin{equation}
	U_G(\tau^k \, \id)U_G^\dagger = \tau^k \, \id,
\end{equation}
the Clifford group acts on $H(N)/Z(N)\simeq \ZZ_N^2$. It is easy to
see that the action of the Clifford group on $H(N)/Z(N)$ is precisely
isomorphic to $SL(2,N)$. This equivalence holds irrespective of
whether $N$ is even or odd and delivers the converse statement
promised above. A proof is given in Appendix A.

%
%
%
%

Symplectic matrices of order 3 are of special interest. If $N > 3$ it can be 
shown that a symplectic matrix is of order 3 if and only if its trace equals $-1 \mod N$ \cite{Marcus}. According to a precise form 
of Zauner's conjecture a SIC fiducial can always be chosen to be 
an eigenvector of the unitary $U_{\cal Z}$ representing the symplectic matrix 

\begin{equation} G_{\cal Z} = \left( \begin{array}{cc} 0 & - 1 \\ 
1 & -1 \end{array} \right) \ . \end{equation}

\noindent In the standard representation $U_{\cal Z}$ is not given in 
monomial form (although monomial representations of other symplectic 
matrices of order 3 can be found for special values of $N$ 
\cite{Marcus}). 

The phase of $U_{\cal Z}$ is chosen so that $U_{\cal Z}^3$ is the identity 
operator. It is then possible, using Gauss sums, to deduce the 
dimension of the three eigenspaces of a Zauner unitary in arbitrary 
dimensions \cite{Zauner}. There are three eigenspaces ${\cal E}_0, 
{\cal E}_1, {\cal E}_2$ corresponding respectively to the eigenvalues 
$1, e^{2\pi i/3}, e^{4\pi i/3}$. With an appropriate choice of phase one 
finds for their dimensions that  

\

\begin{tabular}{|l|l|l|l|} \hline  
$N$ & $1$ & $e^{2\pi i/3}$ & $e^{4\pi i/3}$ \\
\hline 
$3k$ & $k+1$ & $k$ & $k-1$ \\
$3k+1$ & $k+1$ & $k$ & $k$ \\ 
$3k+2$ & $k+1$ & $k+1$ & $k$ \\
\hline 
\end{tabular} 

\

\

\noindent The numerical evidence \cite{Scott} strongly suggests that 
the eigenspace ${\cal E}_0$ always contains SIC fiducials, while if 
$N = 3k$ or $N = 3k+1$ the other two eigenspaces never do. The case 
$N = 3k+2$ does not occur if $N = n^2$. 

Finally we will be interested in the extended Clifford group, which 
includes anti-unitary operators as well \cite{Marcus}. The determinants of the 
2 by 2 matrices are then allowed to take the values $\pm 1$. 
The extended Clifford group divides the set of all SICs into orbits 
in a natural way: acting on a given SIC with an element of this group 
produces another SIC since we assume that the SIC itself is an orbit under 
the Weyl-Heisenberg group.   

\vspace{10mm}

{\bf 3. Representation with phase-permutation matrices}

\vspace{5mm}

\noindent From now on we are in a Hilbert space of dimension $N = n^2$, 

\begin{equation} {\cal H}_N = {\cal H}_n\otimes {\cal H}_n \ . \end{equation}

\noindent We add one more phase factor to the ones we have defined 
already:

\begin{equation} \omega = e^{\frac{2\pi i}{N}} \ , \hspace{8mm} 
\tau = - e^{\frac{i\pi }{N}} \ , \hspace{8mm} 
\sigma = e^{\frac{2\pi i}{n}} \ . \end{equation} 

\noindent Our key observation is that 

\begin{equation} ZX = \omega XZ \hspace{5mm} \Rightarrow \hspace{5mm} 
[X^n, Z^n] = 0 \ . \end{equation}

\noindent Hence the Weyl-Heisenberg group admits an Abelian 
subgroup, 
\begin{equation}
S=\left\langle X^n,Z^n,\tau \mathbb{I}\right\rangle,
\end{equation} 
of maximal order $N\bar{N}$. By diagonalising the elements of $S$, we define a new basis with basis 
vectors labelled $|r,s\rangle $, where $r$ and $s$ are integers 
modulo $n$. Indeed

\begin{equation} X^n|r,s\rangle = \sigma^r|r,s\rangle \ , 
\hspace{8mm} Z^n|r,s\rangle = \sigma^s|r,s\rangle \ . \end{equation}

\noindent Some phase choices are still to be made. We settled for a choice 
which implies that the generators of the group are represented by 

\begin{equation} X|r,s\rangle = \left\{ \begin{array}{ll} |r,s+1\rangle 
& \mbox{if} \ s+1 \neq 0 \ \mbox{mod} \ n \\
\ \\
\sigma^r|r, 0\rangle & \mbox{if} \ s + 1 = 0 \ \mbox{mod} \ n \end{array} \right. 
\label{WHrep} \end{equation} 

\begin{eqnarray} Z|r,s\rangle = \omega^s |r-1,s\rangle \ .  
\end{eqnarray}

\noindent We refer to this representation as the phase-permutation 
representation. It treats the two generators $X$ and $Z$ in  
an even-handed way, so in a sense we have gone 
``half-way'' to the Fourier basis. The relation to the standard 
Weyl basis is given by 

\begin{equation} |r,s\rangle = \frac{1}{\sqrt{n}}\sum_{t=0}^{n-1}
\omega^{-ntr}|nt+s\rangle \ . \label{unittransf} \end{equation}

\noindent The matrix effecting this transformation is of the 
form $F_n\otimes {\bf 1}$, where $F_n$ is the $n$ by $n$ Fourier 
matrix.  

The entire Weyl-Heisenberg group is represented by phase-permutation
matrices, but this time more is true. In this representation the
entire Clifford group is represented by phase-permutation matrices.
The following armchair argument explains why: In all dimensions, the
Clifford group permutes the various maximal Abelian subgroups of the
Weyl-Heisenberg group.  It also preserves the order of any group
element. But if $N = n^2$ there is a unique maximal Abelian subgroup
whose elements (modulo phases) have orders that are the divisors of $n$; namely the one that defines our
basis. Therefore the Clifford group reorders the elements of this
Abelian subgroup, and it follows that it reorders the basis vectors
while possibly multiplying them with phases. 

The permutations involved are easy to deduce. Let $G$ be a general 
symplectic matrix as given in eq. (\ref{eqn:Gmatrix}). Using its inverse we 
observe that  

\begin{equation} U^{\dagger}_GX^nU_G = U^{\dagger}_GD_{n0}U_G = 
D_{\delta n, -\gamma n} = \tau^{-\gamma \delta N}X^{\delta n}Z^{-\gamma n} 
\ \ \end{equation}

\begin{equation} U^{\dagger}_GZ^nU_G = U^{\dagger}_GD_{0n}U_G = 
D_{- \beta n, \alpha n} = \tau^{-\alpha \beta N}X^{- \beta n}Z^{\alpha n} 
\ . \end{equation}

\noindent As usual the case of odd $n$ is simpler, since $\tau^N = 1$ 
in this case. Let us therefore assume that $n$ is odd to begin with. 
We see that 

\begin{equation} X^nU_G|r,s\rangle = U_GX^{\delta n}Z^{-\gamma n}|r,s\rangle 
= \sigma^{\delta r - \gamma s}U_G|r,s\rangle \ \ \end{equation}

\begin{equation} Z^nU_G|r,s\rangle = U_GX^{-\beta n}Z^{\alpha n}|r,s\rangle 
= \sigma^{-\beta r + \alpha s}U_G|r,s\rangle \ . \end{equation}

\noindent It follows that $U_G|r,s\rangle$ is a common eigenvector of 
the diagonal operators $X^n$ and $Z^n$, and indeed that 

\begin{equation} n \ \mbox{odd} \hspace{5mm} \Rightarrow \hspace{5mm} U_G|r,s\rangle = e^{i\theta_{rs}}|\delta r - \gamma s, 
- \beta r + \alpha s\rangle \ , \label{monomial} \end{equation}

\noindent where $\theta_{rs}$ is a phase to be determined. It again 
follows that an arbitrary symplectic unitary is represented by a 
phase-permutation matrix. 

The argument can be extended to the even dimensional case, but since 
we also need to calculate the phases $\theta_{rs}$ we will proceed a 
little differently in the general case. First we define 

\begin{equation} m = \left\{ \begin{array}{lll} 0 & & n \ \mbox{is odd} \\ 
\\ \frac{n}{2} & & n \ \mbox{is even} \ . \end{array} \right. \label{m} \end{equation}

\noindent We may assume 
that the matrix element $\beta$ is relatively prime to $\bar{N}$, 
because if it is not we can fall back on the decomposition 
(\ref{decomp}). Using the standard representation (\ref{standardrep}), 
and relation (\ref{unittransf}) between the two bases, it is 
straightforward to show that 

\begin{eqnarray} \langle r',s'|U_G|r,s\rangle = \frac{e^{i\theta}}
{n^2}\tau^{\beta^{-1}(\delta s'^2 - 2ss'+\alpha s^2)} \times 
\hspace{20mm} \nonumber \\ \ \\ 
\hspace{14mm} 
\times \sum_{t,t'=0}^{n-1}\omega^{nt(-r+\beta^{-1}(-s' + \alpha s + 
m\alpha))}\omega^{nt'(r' + \beta^{-1}(\delta s' - s + m\delta))} \ , 
\nonumber \end{eqnarray} 

\noindent where $m$ was defined in eq. (\ref{m}). Performing the 
sums, and using the fact that $\alpha \delta - \beta \gamma = 1$ 
modulo $n$, we see that 

\begin{equation} \langle r',s'|U_G|r,s\rangle = e^{i\theta}
\tau^{\beta^{-1}(\delta s'^2 - 2ss' + \alpha s^2)} \hspace{2mm} 
\Leftrightarrow \hspace{2mm} \left\{ \begin{array}{l} r' = 
\delta r - \gamma s - m\frac{\delta (1+\alpha )}{\beta} \\ \\
s' = - \beta r + \alpha s + m\alpha \hspace{5mm} , \end{array} 
\right. \end{equation}  

\noindent and zero otherwise. If $n$ is odd then $m=0$, and 
we have reproduced eq. (\ref{monomial}) but with the phases 
now included. If $n$ is even we use modulo 2 arithmetic 
to polish the $m$-dependent term; note that $\beta = 1$ modulo 
2 since $\beta$ is relatively prime to $n$. Thus we arrive at 
our key result:

\

\noindent {\bf Theorem 1}: {\sl When the dimension is a square 
number $N = n^2$, the Clifford group admits a representation by 
phase-permutation matrices. For the Weyl-Heisenberg subgroup it 
is given by eqs. (\ref{WHrep}). For an $SL(2,\bar{N})$ element of the 
form (\ref{eqn:Gmatrix}), with $\beta$ and $N$ relatively prime and 
$m$ and $s'$ as above, it is} 

\begin{equation} U_G|r,s\rangle = e^{i\theta}\tau^{\beta^{-1}
(\delta s'^2 -2ss' + \alpha s^2)}|\delta r - \gamma s + 
m \gamma \delta, - \beta r + \alpha s + m\alpha \beta \rangle 
\ . \label{therep} \end{equation}

\noindent {\sl The overall phase $\theta$ remains undetermined. The 
case when $\beta$ is not relatively prime to $N$ can be recovered 
from eq. (\ref{decomp}).} 

\bigskip

\noindent Note that if $G = G'$ modulo $n$ (using modulo $n$ arithmetic for 
the matrix elements) then 
$U_G$ and $U_{G'}$ produce the same permutations of the basis 
elements, that is to say they differ only by a diagonal unitary.

The group element of most interest to us is Zauner's unitary, corresponding 
to the matrix (\ref{ZaunerG}). It is given explicitly by 

\begin{equation} U_{\cal Z}|r,s\rangle = e^{\frac{i\pi(N-1)}{12}}\tau^{r^2 + 2rs}
|-r-s-m,r\rangle \ . \label{ZaunerU} \end{equation}

\noindent Here the overall phase $\theta$ was chosen to ensure that 
$U_{\cal Z}^3$ is the identity operator.  

A general element of the extended Clifford group is obtained 
by replacing $SL(2,\bar{N})$ with the group $ESL(2,\bar{N})$. The additional matrices $E \in ESL(2,\bar{N})$ can be written as a product  

\begin{equation} E = GJ \ , \hspace{6mm} G \in SL(2,\bar{N}), 
\hspace{6mm} J = \left( \begin{array}{cc} 1 & 0 \\ 0 & -1 
\end{array} \right) \ . \end{equation}

\noindent To the matrix $J$ there corresponds an anti-unitary 
operator $U_J$ whose action on the phase-permutation basis is 
given by 

\begin{equation} U_J|r,s\rangle = |-r,s\rangle \ . \end{equation}

\noindent Hence, the extended Clifford group also acts 
through phase-permutation matrices on the basis vectors. 

\vspace{10mm}

{\bf 4. Uniqueness proofs}

\vspace{5mm}

\noindent We have seen by construction that the Clifford group admits
a representation by phase-permutation matrices if the dimension $N =
n^2$ is a square. We will now prove the converse, that this is
possible only in square dimensions. Since the construction hinged on a
special maximal Abelian subgroup of the Weyl-Heisenberg group---it
transformed into itself under the Clifford group---we begin with a
theorem that shows that this is a necessary feature of any
phase-permutation representation. It actually applies to a slightly
more general situation, in which we consider a group ${\cal G}$ which
may be the entire Clifford group, but which may also be any subgroup
of the Clifford group which includes the Heisenberg group $H(N)$ as a
subgroup.

An Abelian subgroup $S$ of $H(N)$ is called
\emph{maximal} if no element $g\in H(N)\setminus S$ commutes with everything in $S$. Equivalently, $S$ is maximal if it has the maximal possible order $N \bar{N}$.
\

\noindent {\bf Theorem 2}: {\sl The following two statements are equivalent:} 

\

{\sl 1. There exists a phase-permutation representation of ${\cal G}$ on 
$\CC^N$ which is 

irreducible when restricted to $H(N)$.} 

\

{\sl 2. There exists a maximal Abelian subgroup of $H(N)$ which is stabilised 

by ${\cal G}$.}

\

\noindent {\sl Proof}: 
First we establish $2 \Rightarrow 1$.
Let $S$ be a maximal Abelian subgroup of $H(N)$. Choose a basis in
which all elements of $S$ are simultaneously diagonal. If we select
one representative from every coset in $S/Z(N)$, their diagonals
define $N$ orthogonal vectors (because $H(N)/Z(N)$ defines a unitary
operator basis). From this one concludes that a maximal Abelian
subgroup defines a joint eigenbasis which is unique up to permutations
and rephasings.  It is then obvious that $2 \Rightarrow 1$. This was
used in section 3.
 
To prove that $1 \Rightarrow 2$, denote the phase-permutation basis by
$\{ |e_a\rangle \}_{a=1}^N$. By assumption $H(N)$ simply permutes the
corresponding rays (vectors up to phase), and acts transitively on
them. Let $S$ be the subgroup of the Heisenberg group leaving the
particular projector $|e_1\rangle \langle e_1|$ invariant. Since the
orbit of $H(N)$ acting on the set of projectors has size $N$ we know
that $|H(N)|/|S| = N$, that is to say that  

\begin{equation} |S| = \frac{|H(N)|}{N} = \frac{N^2\bar{N}}{N} = N\bar{N} 
\ , \end{equation}

\noindent where $\bar{N}$ is the cardinality of the center. Modulo phases, 
$S$ must thus be a subgroup of $H(N)$ of order $N$.  By definition,
all elements of $S$ have the common eigenvector $|e_1\rangle $.  But
it is a direct consequence of the commutation relations that two
elements of the Heisenberg group have a common eigenvector if and only
if they commute.  Therefore $S$ is an abelian subgroup of $H(N)$ and
has order $N\bar{N}$.  Hence $S$ is maximally Abelian.

Now $H(N)$ acts monomially on the joint eigenvectors of any maximal abelian 
subgroup. It also acts transitively, by irreducibility. Thus the orbit of 
$|e_1\rangle $ under $H(N)$ consists precisely of the joint eigenvectors of 
$S$ (up to phases). But we showed before that the orbit coincides with 
$\{ |e_a\rangle \}_{a=1}^N$.  $\Box$

We now focus on the case where ${\cal G}$ is the full Clifford group.
Are there cases beyond square dimensions where there is a maximally
Abelian subgroup $S$ of $H(N)$ stabilized by the Clifford group? By
Section~2, the orbit of an element of $H(N)/Z(N)$ under the action of
the Clifford group corresponds to the orbit of a vector, $v\in\ZZ_N^2$,
under the action of $SL(2,N)$.
The following characterization of these orbits is implied by Lemma~27
of \cite{Gross}. We re-prove it here to make the
presentation self-contained.

Recall that the \emph{order} of a vector $v\in\ZZ_N^2$ is the least
integer $k\geq 1$ such that $k v \equiv 0$, where the triple bar
equality sign denotes equality modulo $N$.

\ 

\noindent {\bf Lemma 1}: 
{\sl 
	The orbits of the action of $SL(2,N)$ on $\ZZ_N^2$
	are the sets 
	\begin{equation}
		{\mathcal O}_k = \{ v\in\ZZ_N^2 \,|\, \ord v = k \}
	\end{equation}
	of vectors of constant order.
}

\ 

\noindent {\sl Proof}: 
	We first prove that $S\in SL(2,N)$ cannot change the
	order of a vector $v$. Indeed, if $k v \equiv 0$, then
	\begin{equation}
		k (S v) \equiv S( k v) \equiv S 0 = 0
	\end{equation}
	so that $\ord S v \leq \ord v$. Replacing $S$ by $S^{-1}$, we see
	that $\ord S v \geq \ord v$. Hence equality must hold as claimed.

	The comparatively difficult part is to show that
	$SL(2,N)$ acts transitively on the sets
	${\mathcal O}_k$. Let $v\in\ZZ_N^2$ be of order $k$. By 
	definition we thus have that
	\begin{equation}
		N\,|\,(k v_i) \Rightarrow \frac Nk \,|\, v_i,
	\end{equation}
	where $v_i, i = 1, 2$ are the components of $v$. Therefore, the
	vector
	\begin{equation}
		v' := \frac{1}{N/k} v
	\end{equation}
	is well-defined as an element of $\ZZ_N^2$. One checks that $\ord v'
	= N$.

	We go on to show that there is a symplectic matrix
	$S\in SL(2,N)$ whose first column equals $v'$. That is
	the case if there are integers $x,y$ such that
	\begin{equation}\label{eqn:det}
		1\equiv\det
		\left(
			\begin{array}{cc}
				v_1' & x \\
				v_2' & y 
			\end{array}
		\right)
		=v_1'y - v_2' x.
	\end{equation}
	By B\' ezout's identity, there are integers $a,b$ such that
	\begin{equation}
		v_1' a + v_2' b = g,
	\end{equation}
	where $g={\mathrm gcd}(v_1',v_2')$. It must be the case that $g$ and
	$N$ are co-prime, for otherwise $(N/g)$ would be an integer smaller
	than $N$ such that $(N/g) v' \equiv 0$, which would contradict the
	fact that $\ord v' =N$. Thus there exists a multiplicative inverse
	$g^{-1}$ of $g$ modulo $N$. Hence 
	\begin{equation}
		y=g^{-1} a,\qquad x = - g^{-1} b
	\end{equation}
	provides a solution to (\ref{eqn:det}).

	Finally, let $w$ be another vector of order $k$. Let $S_v$ be a
	symplectic matrix with first column equal to $v'$, let $S_w$ be a
	symplectic matrix with first column equal to $w'$. Then
	\begin{equation}
		S_w S_v^{-1} v' \equiv w' \Rightarrow S_w S_v^{-1} v \equiv w.
	\end{equation}	
	Thus any two elements of ${\mathcal O}_k$ can be mapped onto each
	other by means of an element of $SL(2,N)$.
	$\Box$

The preceding lemma allows us to decide in which dimensions there is a
monomial representation of the Clifford group just by counting orbit
sizes. It seems simpler to do that in prime-power dimensions.

\ 

\noindent {\bf Lemma 2}: 
{\sl
	Let $N=p_1^{q_1} \dots p_k^{q_k}$ be the decomposition of the
	dimension into powers of distinct primes. 
	
	There is an order-$N$ subgroup of $\ZZ_N^2$ which is stabilized by
	$SL(2,N)$ if and only if the same is true for all dimensions $N_i =
	p_i^{q_i}$, for $i=1, \dots, k$.
}

\ 

The statement follows from the more general fact that the
Weyl-Heisenberg group and the Clifford group factor into direct
products for composite $N$. It is proven in Appendix B. We are
ready to conclude this section:

\ 

\noindent {\bf Theorem 3}:
{\sl
	There exists a monomial
	representation of the Clifford group which contains the
	Weyl-Heisenberg group as an irreducible subgroup if and only if the
	dimension $N=n^2$ is a square.
}

 \

\noindent{\sl Proof}:
	Using the notions of Lemma 2, 
	let $N_i =
	p_i^{q_i}$. Let $V\subset \ZZ_{N_i}^2$ be a non-trivial subgroup which is
	invariant under the action of $SL(2,N)$.
	
	Let $k=\max\{\ord v\,|\,v\in V\}$ be the largest order of any
	element in $V$. By Lagrange's Theorem, $k$ is of the form $k=p_i^l$
	for $0\leq l \leq q_i$. Because $v\in V \Rightarrow p_i v \in V$,
	there is also an element of order $p_i^{l-1}$ in $V$, and, indeed, any
	power of $p_i$ up to the $l$th appears as the order of some element
	in $V$.

	By Lemma 1 
	and the assumption that $V$ be
	$\mathrm{SL}$-invariant, we find that
	\begin{equation}
		V=\{ v \in \ZZ_{N_i}^2 \,|\, \ord v \leq p_i^l\} = p_i^{q_i -l
		}\,\ZZ_{N_i}^2.
	\end{equation}
	Hence $|V| = p_i^{2l}$, which is equal to $N_i$ if and only if $l =
	q_i/2$. That is possible if and only if $q_i$ is even, which implies
	the claim.
$\Box$

In the remaining sections we turn our attention to applications of the
phase-permutation basis. We return to our original motivation of
SICs. First we discuss a general property of SICs on the phase-permutation basis, and then in sections 6 and 7 we use this basis to construct SICs in dimensions $2^2$, $3^2$, and $4^2$. The sixteen dimensional case has so far not been solved in the Weyl basis.  Finally, in section 8 we consider sets of Mutually Unbiased Bases (MUB) in the phase-permutation basis.

\vspace{10mm}

{\bf 5. Images of SICs in the probability simplex}

\vspace{5mm}

\noindent For the first application, recall that a quantum state together 
with an orthonormal basis gives rise to a probability vector, with the 
basis vectors themselves forming a simplex. In 
physics this is connected to quantum measurements, in mathematics it is 
an example of the moment map. Anyway, if the quantum state is a pure 
state represented by the Hilbert space vector  

\begin{equation} (z_0, z_1, \dots , z_{N-1})^{\rm T} = 
(\sqrt{p_0}, \sqrt{p_1}e^{i\mu_1}, \dots , 
\sqrt{p_{N-1}}e^{i \mu_{N-1}})^{\rm T} \ , \label{komponenter} \end{equation}

\noindent then its image with respect to the same orthonormal basis is
a probability vector with components $p_i$.  The probability vector
gives the barycentric coordinates of a point within the probability
simplex associated to the basis chosen. We assumed that the vector is
a unit vector, which indeed implies that

\begin{equation} \sum_{i = 0}^{N-1} p_i = 1 \ . \end{equation}

\noindent If we have a set of such pure states connected by a group
which is represented by phase permutation matrices, then their
probability vectors will be related by permutations of the coordinate axes. It follows that they lie on a sphere centered around the
midpoint of the probability simplex.

More can be said if we deal with the $N^2$ unit vectors forming an
orbit under the Weyl-Heisenberg group and obeying the SIC condition.
Let the components of a SIC fiducial vector be given by eq.
(\ref{komponenter}). In the standard basis, where the subgroup generated by $Z$ is diagonalised, it is known that \cite{Mahdad, Chris}

\begin{equation} \sum_{i=0}^{N-1} p_ip_{i+x} = \left\{ \begin{array}{lll} \frac{2}{N+1} & & \mbox{if} \ x = 0 \\ \ \\ 
\frac{1}{N+1} & & \mbox{if} \ x \neq 0 \ \mbox{mod} \ N \ . 
\end{array} \right. \label{ADF} \end{equation}

\noindent The length of the probability vector, that is the radius of 
the sphere in which it is inscribed, is thereby determined. The remaining 
equations also admit an interesting geometrical interpretation \cite{Appleby}.
Since the matrix representing $Z$ is diagonal, the $N$ states $Z^r|\psi_F\rangle $ 
all project to the same probability vector. Moreover the $i$th component of the 
vector $X^x|\psi_F\rangle $ is $p_{i-x}$, so from these equations we can read 
off not only the length of the probability vectors but also their mutual angles. 
The conclusion turns out to be that the image of the $N^2$ SIC vectors is itself 
a regular simplex with $N$ vertices. Once this geometrical interpretation 
is available it is unsurprising 
that the $N$ equations (\ref{ADF}) are redundant and do not 
by themselves determine all the coefficients $p_i$ (unless 
$N = 2$). But they are still helpful. 

The reason why the $N^2$ vectors in the SIC give rise to only 
$N$ images in the projection is that the images form an orbit 
under the subgroup that is complementary to the diagonalised 
subgroup. It is clear that something similar should happen in the 
phase-permutation basis, where again there is an Abelian subgroup 
of order $N$ that does not move the projected points. To show 
this we denote the components of the SIC fiducial $|\psi_F\rangle$ by 
$z_{rs} = \sqrt{p_{rs}}e^{i\mu_{rs}}$, and find  

\begin{eqnarray} \langle \psi_0|X^{nu}Z^{nv}|\psi_0\rangle = 
\sum_{r,s=0}^{n-1}p_{rs}\sigma^{ru+sv} \hspace{20mm} \nonumber \\ 
\Rightarrow |\langle \psi_0|X^{nu}Z^{nv}|\psi_0\rangle |^2 = \sum_{r,s=0}^{n-1} \sum_{r',s'= 0}^{n-1} 
p_{rs}p_{r's'}\sigma^{(r-r')u + (s-s')v} \ . \label{MUSaux} \end{eqnarray}

\noindent The absolute values on the left hand side are known from the 
condition defining a SIC. Using this, and summing over the integers $u$ and $v$, we find 

\begin{equation} \sum_{r,s=0}^{n-1}p^2_{rs} = \frac{1}{N}\left( 1 + \frac{N-1}{N+1}
\right) = \frac{2}{N+1} \ . \label{MUS} \end{equation}

\noindent More generally we can take a Fourier transform of eq. 
(\ref{MUSaux}). This gives $N-1$ additional equations for the 
absolute values, 

\begin{equation} \sum_{r,s=0}^{n-1}p_{rs}p_{r+x,s+y} = \frac{1}{N+1} 
\ . \label{moduli} \end{equation}

\begin{figure}[http] 

       \hbox{
              \epsfig{figure=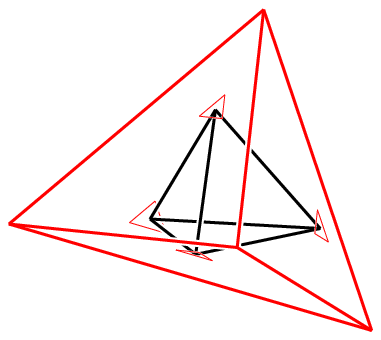,width=.41\linewidth}
       \hspace{8mm}
               \epsfig{figure=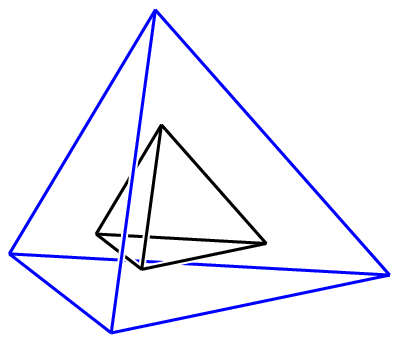,width=.41\linewidth}}
       \caption{\small The images of an $N = 4$ SIC, in the two 
       bases we discuss. The phase-permutation basis (the large 
       tetrahedron on the right) is aligned to the SIC (the small 
       tetrahedron) in a way that the standard basis (the large 
       tetrahedon on the left) 
       is not. The 
       corresponding pictures for $N = 9$ (not shown) are eight 
       dimensional, and would show that the faces of the eigenvalue 
       simplex for the phase-permutation basis are nicely aligned 
       with the image of the SIC.}
       \label{fig:SIC4-1}
\end{figure}

\noindent Here $x,y$ are integers modulo $n$, not both zero. 
This is analogous to what happens in the standard basis, and 
the geometrical interpretation is the same: when the SIC is 
projected to the basis simplex we see a regular simplex centered 
at the origin with just $N$ vertices. Its orientation 
differs from the projection to the standard 
basis (see Fig \ref{fig:SIC4-1}). 

\vspace{10mm}

{\bf 6. SIC fiducials for $N=2^2$ and $3^2$}

\vspace{5mm}

\noindent If we use the phase-permutation representation it 
becomes very easy to find the SICs in $N = 2^2$ dimensions. In 
fact the absolute values entering the fiducial are determined 
by Zauner's conjecture, normalisation, and eq. (\ref{MUS}). 
Alternatively one can simply solve the equations defining a 
Weyl-Heisenberg covariant SIC. Before doing so it is convenient 
to rephase the basis through

\begin{equation} (|0,0\rangle , \ |0,1\rangle , \ |1,0\rangle , \ |1,1\rangle ) 
\ \rightarrow \ (\tau^{-2}|0,0\rangle , \ \tau^{-7}|0,1\rangle , \ 
\tau^{-5}|1,0\rangle , \ |1,1\rangle ) \ . \end{equation}

\noindent This gives the representation  

\begin{equation} X = \tau \left( 
\begin{array}{cccc} 0 & i & 0 & 0 \\ -1 & 0 & 0 & 0 \\ 
0 & 0 & 0 & 1 \\ 0 & 0 & i & 0 \end{array} 
\right) \hspace{8mm} Z = \tau \left( 
\begin{array}{cccc} 0 & 0 & -1 & 0 \\ 0 & 0 & 0 & 1 \\ 
i & 0 & 0 & 0 \\ 0 & i & 0 & 0 \end{array} 
\right) \ . \end{equation}

\noindent There are altogether $4^4 = 256$ possible solutions for a SIC 
fiducial, giving rise to 16 SICs altogether. The solutions are 

\begin{equation} \left( \begin{array}{c} x \\ i^{s_1} \\ i^{t_1} 
\\ i^{u_1} \end{array} \right) \ \ \left( \begin{array}{c} i^{s_2} \\ x \\ i^{t_2} 
\\ i^{u_2} \end{array} \right) \ \ \left( \begin{array}{c} i^{s_3} \\ i^{t_3} \\ x 
\\ i^{u_3} \end{array} \right) \ \ \left( \begin{array}{c} i^{s_4} \\ i^{t_4} 
\\ i^{u_4} \\ x \end{array} \right) \ , \end{equation}

\noindent where $s_i,t_i,u_i$ are integers from 0 to 3, 

\begin{equation} x = \sqrt{2+\sqrt{5}} \ , \end{equation}

\noindent and an overall normalisation has been ignored. 
This simple form of the SIC 
vectors has been found before by Zauner \cite{Zauner2}, who arrived at it by casting the Zauner unitary into phase-permutation form, and from an alternative point of view by Belovs \cite{Belovs}. 
Each of the vectors is left invariant by a Zauner unitary of order 3, and 
the 16 SICs form a single orbit under the Clifford group. The result of the 
numerical searches \cite{Renes, Scott} is thus fully confirmed. The images 
in the probability simplex (see section 5, and Fig. \ref{fig:SIC4-1}) coincide 
for all the SICs, and they are nicely 
oriented. The structure of the $N = 4$ SICs, and their entanglement 
properties, were studied in detail recently \cite{Zhu}.  

The case of $N = 3^2$ is much harder. Lest our readers be disappointed
by this, we strongly recommend they begin by looking at the answer
found---by means of a Magma calculation---in the standard
representation \cite{Scott, Grassl}. Afterwards our result will come
as a pleasant surprise.

With our canonical choice of $U_{\cal Z}$, Zauner's conjecture 
implies that the SIC fiducial takes the special form

\begin{eqnarray} |\psi \rangle = - z_1 \omega^7|1,1\rangle - z_2 \omega |2,2 \rangle 
+ \hspace{60mm} \nonumber \\ \ \\ 
+ z_3 (\omega^6|0,2\rangle + |1,0\rangle + \omega^8|2,1\rangle ) + 
z_4(\omega^6|0,1\rangle + |2,0\rangle+ \omega^5|1,2\rangle  ) \ . 
\label{fiducial}
\nonumber \end{eqnarray}

\noindent We included some convenient phase factors. We exploit the arbitrariness in the overall phase to write the $z_j$  in the form

\begin{equation} z_1 = \sqrt{p_1}e^{i\mu_0} \hspace{5mm} 
z_2 = \sqrt{p_2}e^{-i\mu_0} \hspace{5mm} z_3 = \sqrt{p_3}e^{i\mu_3} \hspace{5mm} 
z_4 = \sqrt{p_4}e^{i\mu_4}  \end{equation}

\noindent
where it is assumed that $-\pi/2 < \mu_0 \le \pi/2$.  

The necessary and sufficient condition for a normalized vector $|\psi\rangle$ to be a fiducial vector is that

\begin{equation}
\left|\langle \psi | D_{jk} |\psi\rangle \right|^2 = \frac{1}{10} 
\label{eq:fidCond}
\end{equation}

\noindent for all $j,k$ not both zero---a total of $80$ equations.  However the Zauner symmetry means that if eq.~(\ref{eq:fidCond}) is satisfied for the vector $(j,k)$ it is automatically satisfied for the (up to) two other vectors obtained by acting with the Zauner matrix. Also, if it is satisfied for $(j,k)$ it is automatically satisfied for $(-j,-k)$.  Consequently we can reduce the $80$ equations to $15$, which it will be convenient to group as follows
\begin{enumerate}
\item $2$ group $1$ equations

\begin{equation}
\left|\langle \psi | D_{0,3} |\psi\rangle \right|^2 = \left|\langle \psi | D_{3,6} |\psi\rangle \right|^2=\frac{1}{10} 
\end{equation}

\noindent  Assuming normalization these equations  are equivalent to eqs.~(\ref{moduli}).
\item $3$ group $2$ equations

\begin{equation}
\left|\langle\psi |D_{3j+1,6j+2} |\psi\rangle\right|^2 = \frac{1}{10}
\end{equation}

\noindent with $j=0,1,2$.
\item $9$ group $3$ equations

\begin{equation}
\left| \langle \psi | D_{3j,6j+3k+1} | \psi\rangle\right|^2 =\frac{1}{10}
\end{equation}

\noindent with $j,k=0,1,2$.
\end{enumerate}
Together with normalization this gives us $16$ conditions on the $7$ real parameters in eq.~(\ref{fiducial}), so there is still a high degree of redundancy in the equations.  The nature of the dependencies will become clear in the course of solving them.

We begin by considering the normalization condition and  group $1$ equations.  As can be seen from   eqs.~(\ref{MUS}) and~(\ref{moduli}) (which are equivalent to them) they only involve the absolute values.  Explicitly:

\begin{equation}  \begin{array}{rl} 
p_1+p_2+3p_3+3p_4 &= 1
\\
\
\\
p_1^2 + p_2^2 - p_1p_2 &= 
\frac{1}{10} \\ \ \\ 
3p_3^2 + 3p_4^2 + 3p_3p_4 - p_3 - p_4 & = 
- \frac{1}{10} \end{array}  \ . \label{9eq} \end{equation}  

\noindent These equations are not hard to solve. Setting

\begin{equation} 
p_1 = a_1+b_1 \ , \hspace{5mm} p_2 = a_1 - b_1 \ , 
\hspace{8mm} p_3 = a_3 + b_3 \ , \hspace{5mm} p_4 = a_3 - b_3 
\end{equation}

\noindent diagonalizes them. It is then readily deduced

\begin{equation}
a_3  = \frac{1}{6} (1-2a_1)
\ , \hspace{5 mm}
b_1^2  = \frac{1}{30} (1-10 a_1^2)
\ , \hspace{5 mm}
b_3^2  = \frac{1}{180}(-1+20 a_1-60 a_1^2) \  .
\label{a3b1b2Sols}
\end{equation}

To fix the free parameter in the expressions just derived we need to consider the group $2$ equations.  This will also give us the phase $e^{i\mu_0}$.  It is convenient to write the equations in the form

\begin{equation}
\sum_{r,s=0}^2 \sigma^{j(r-s)} e^{\vphantom{*}}_r e^{*}_s= \frac{1}{10}
\label{gp2B}
\end{equation}

\noindent with $j=0,1,2$ and where

\begin{equation}
e_0 = z^{\vphantom{*}}_1 z^{*}_2\ , \hspace{20 mm} 
e_1 = i\sqrt{3}\omega^2 p_3 \ , \hspace{20 mm}
e_2 = -i \sqrt{3}\omega^7 p_4 \ .
\end{equation}

\noindent Inverting the Fourier transform we see that eqs.~(\ref{gp2B}) are equivalent to the $2$ conditions

\begin{eqnarray}
|e_0|^2 + |e_1|^2 + |e_2|^2  = \frac{1}{10} \nonumber
\\
\
\\
e^{\vphantom{*}}_0 e^{*}_1 + e^{\vphantom{*}}_1 e^{*}_2 +e^{\vphantom{*}}_2 e^{*}_0 = 0 \nonumber
\end{eqnarray}

\noindent  The first of these is a consequence of the group $1$ equations.  Solving the second for $e^{\vphantom{*}}_0 = z^{\vphantom{*}}_1z^{*}_2$ gives

\begin{equation}
z^{\vphantom{*}}_1 z^{*}_2 =\frac{e^2_2e^{*}_1 - e^2_1e^{*}_2}{|e_1|^2 - |e_2|^2}= \frac{\sqrt{3}(\vphantom{\sqrt{3}}a_3^2-b_3^2)(\sqrt{3}b_3-ia_3)}{4 a_3 b_3} \ . \label{z1z2s}
\end{equation}

\noindent Taking the square of the absolute value on both sides and using eqs.~(\ref{a3b1b2Sols}) we find

\begin{equation}
(1-40a_1+40a_1^2)(-11+100a_1-120a_1^2-800a_1^3+1600a_1^4) = 0 \ .
\end{equation}

\noindent Solving this equation and taking account of the requirement that $b_1$, $b_3$  both be real we deduce

\begin{eqnarray} a_1 = \frac{1}{40}\left(5 - s_05\sqrt{3} + 
s_03\sqrt{5} + \sqrt{15}\right) \hspace{11mm} \nonumber \\
\nonumber \\
b_1 = \frac{s_2}{60}\sqrt{15\left(\sqrt{15}+s_0\sqrt{3}\right)} \hspace{30mm}
\nonumber \\  \label{absval9} \\ 
a_3 = \frac{1}{120}\left(15 + s_05\sqrt{3} - s_03\sqrt{5} - \sqrt{15}\right) 
\hspace{8mm} \nonumber \\ 
\nonumber \\ 
b_3 = \frac{s_1}{60}\sqrt{5\left(-18 - s_07\sqrt{3} + s_06\sqrt{5} + 5\sqrt{15}\right)}
\nonumber  \end{eqnarray}

\noindent where $s_0$, $s_1$ and $s_2$ are arbitrary signs. This fixes the absolute values.  Note that the only choice of sign that affects the set of absolute values is $s_0$, 
which suggests---correctly---that $s_0$ labels two different orbits of the Clifford 
group. 

To determine the phase $e^{i\mu_0}$, we substitute these expressions into eq.~(\ref{z1z2s}) and simplify.  We obtain

\begin{equation}
e^{2 i \mu_0}
=
\frac{1}{4} \sqrt{2\left(6+s_0\sqrt{3} -\sqrt{15} \right)} -\frac{i s_1}{4}\sqrt{2\left(2-s_0\sqrt{3} +\sqrt{15} \right)}
\end{equation}

\noindent Taking account of the assumption that $-\pi/2 < \mu_0 \le \pi/2$ we deduce
\begin{equation} e^{i\mu_0} = \sqrt{\frac{1}{2}+c_0}-is_1\sqrt{\frac{1}{2}-c_0}  
\end{equation}
\noindent where
\begin{equation}
c_0 = \frac{1}{8}\sqrt{2(6+s_0\sqrt{3}-\sqrt{15})} \ . 
\end{equation}

\noindent Note that the numbers given in these expressions as nested square roots can  be constructed with ruler and compass, so the ancient Greeks might have 
approved---especially since the 9th root of unity cannot be so constructed.

To calculate the remaining two phases we turn to the group $3$ equations.  It is convenient to write the equations in the form

\begin{equation}
\sum_{r,s=0}^{2} \sigma^{j(r-s)}e^{\vphantom{*}}_{kr}e^{*}_{ks} = \frac{1}{10}
\label{gp3B}
\end{equation}

\noindent with $j,k=0,1,2$ and where

\begin{equation}
\begin{array}{rl}
e_{k0} &= \left(1+2(-1)^k \cos \frac{(3k+2)\pi}{9}  \right)z^{*}_3 z^{\vphantom{*}}_4 \nonumber
\\
&
\\
e_{k1}& = - \left(\tau^{6k-5} z^{*}_1 + \tau^{-(6k-5)} z^{*}_2  \right) z^{\vphantom{*}}_3
\\
&
\\
e_{k2} & = -\left(  \tau^{6k-5} z^{\vphantom{*}}_1 +\tau^{-(6k-5)} z^{\vphantom{*}}_2\right) z^{*}_4
\nonumber \ .
\end{array}
\end{equation}

\noindent Inverting the Fourier transform in eqs.~(\ref{gp3B}) we see that the nine group $3$ equations are actually equivalent to the six equations 

\begin{eqnarray}
|e_{k0}|^2 + |e_{k1}|^2 + |e_{k2}|^2 &= \frac{1}{10}
\label{gp3C}
\\
\ & \nonumber 
\\
e^{\vphantom{*}}_{k0}e^{*}_{k1}+e^{\vphantom{*}}_{k1}e^{*}_{k2} + e^{\vphantom{*}}_{k2}e^{*}_{k0} &= 0 
\label{gp3D}
\end{eqnarray}

\noindent with $k=0,1,2$.  
Using eq.~(\ref{z1z2s}) and some elementary trigonometry one finds that for all three values of $k$ eq.~(\ref{gp3C}) is equivalent to the single condition

\begin{equation}
3(a_3^2-b_3^2)+4a_1 a_3 = \frac{1}{10}
\end{equation}

\noindent which is an immediate consequence of the group $1$ equations. We are thus left with the three eqs.~(\ref{gp3D}).  It will be convenient to write them in the form 

\begin{equation}
f_1 = f_2 = f_3 = 0
\end{equation}

\noindent where

\begin{equation}
f_j = \sum_{k=0}^{2} \sigma^{jk} \left( e^{\vphantom{*}}_{k0}e^{*}_{k1}+e^{\vphantom{*}}_{k1}e^{*}_{k2} + e^{\vphantom{*}}_{k2}e^{*}_{k0}\right) \ .
\end{equation}

\noindent Writing the expressions out in full we find that $f_2 = \tau^2 f_1$.  So the nine equations with which we started  reduce to just the two equations $f_0=f_1=0$.  It is readily confirmed that these are equivalent to

\begin{eqnarray}
e^{3 i \mu_3} = - \frac{1}{2\sqrt{p_3}}
\left(\frac{z^2_1+z^{\vphantom{2}}_1z^{\vphantom{2}}_2 +z^2_2}{z^{*}_1+z^{*}_2} +\frac{i\left(z^2_1-z^{\vphantom{2}}_1z^{\vphantom{2}}_2 +z^2_2\right)}{\sqrt{3}\left(z^{*}_1-z^{*}_2\right)}
\right) \nonumber
\\
\
\\
e^{3 i \mu_4} = - \frac{1}{2\sqrt{p_4}}
\left(\frac{z^2_1+z^{\vphantom{2}}_1z^{\vphantom{2}}_2 +z^2_2}{z^{*}_1+z^{*}_2} -\frac{i\left(z^2_1-z^{\vphantom{2}}_1z^{\vphantom{2}}_2 +z^2_2\right)}{\sqrt{3}\left(z^{*}_1-z^{*}_2\right)}
\right) \nonumber
\end{eqnarray}

\noindent The quantities on the right hand sides  are all known so these formulae give explicit expressions for the two remaining phases.  Simplifying them and taking the cube roots we find

\begin{eqnarray} e^{i\mu_3} = \sigma^{m_3}\left( -\sqrt{\frac{1}{2} - c_1+c_2} + is_1s_2\sqrt{\frac{1}{2} + c_1-c_2}\right)^{\frac{1}{3}} \ \nonumber \\ 
\ \\ 
e^{i\mu_4} = \sigma^{m_4} \left( - \sqrt{\frac{1}{2} - c_1-c_2} + is_1s_2\sqrt{
\frac{1}{2}+c_1+c_2}
\right)^{\frac{1}{3}} \ , \nonumber \end{eqnarray}

\noindent where 

\begin{eqnarray} c_1 = \frac{s_0}{8}\sqrt{9 - 
s_04\sqrt{3} + s_03\sqrt{5} - 2\sqrt{15}} \hspace{21mm}  \nonumber \\
\ \\ 
c_2 = \frac{s_1s_0}{24}\sqrt{15(-19+s_012\sqrt{3}-s_09\sqrt{5}+6\sqrt{15})} 
\ . \nonumber \end{eqnarray}

\noindent Here $m_3$ and $m_4$ can take the values $0,1,2$. The entire solution 
is given in terms of radicals, as expected (but not understood!). 

Altogether there are $2^3\cdot 3^2 = 72$ fiducial vectors, splitting into 2 different 
orbits of the extended Clifford group labelled by $s_0 = \pm 1$. The solution with $s_0 = s_1 = s_2 = m_3 = m_4 = 1$ is the fiducial 9a as labelled by Scott and Grassl \cite{Scott}, while switching the sign of (only) $s_0$ leads to their fiducial 9b.  


\vspace{10mm}

\newpage

{\bf 7. SIC fiducials for $N=4^2$}

\vspace{5mm}

\noindent The first dimension for which the approach helped in finding a new
solution is $N=16=4^2$.  We were unable to obtain a solution by hand as we did for $N=9$. However, we were able to obtain a solution using Magma.

We use a basis such that both $X^4$ and $Z^4$ are diagonal.  The
change of basis is given by the matrix $T$ in
Figure~\ref{fig:base_change16}, and the Weyl-Heisenberg group in
this basis are shown in Figure~\ref{fig:transformed_XZ}.

\begin{figure}[tb]
\begin{footnotesize}
\begin{alignat*}{5}\let\w\tau\arraycolsep0.5\arraycolsep
T=\frac{1}{2}\left(\begin{array}{*{16}{c}}
1  &0  &0  &0  &\w^{16}&0  &0  &0  &1  &0  &0  &0  &\w^{16}&0  &0  &0  \\
1  &0  &0  &0  &\w^{24}&0  &0  &0  &\w^{16}&0  &0  &0  &\w^{8}&0  &0  &0  \\
\w^{20}&0  &0  &0  &\w^{20}&0  &0  &0  &\w^{20}&0  &0  &0  &\w^{20}&0  &0  &0  \\
1  &0  &0  &0  &\w^{8}&0  &0  &0  &\w^{16}&0  &0  &0  &\w^{24}&0  &0  &0  \\
0  &0  &1  &0  &0  &0  &\w^{16}&0  &0  &0  &1  &0  &0  &0  &\w^{16}&0  \\
0  &0  &1  &0  &0  &0  &\w^{24}&0  &0  &0  &\w^{16}&0  &0  &0  &\w^{8}&0  \\
0  &0  &\w^{8}&0  &0  &0  &\w^{8}&0  &0  &0  &\w^{8}&0  &0  &0  &\w^{8}&0  \\
0  &0  &1  &0  &0  &0  &\w^{8}&0  &0  &0  &\w^{16}&0  &0  &0  &\w^{24}&0  \\
0  &0  &0  &\w^{4}&0  &0  &0  &\w^{20}&0  &0  &0  &\w^{4}&0  &0  &0  &\w^{20}\\
0  &0  &0  &-\w^{19}&0  &0  &0  &-\w^{11}&0  &0  &0  &\w^{3}&0  &0  &0  &-\w^{27}\\
0  &0  &0  &\w^{20}&0  &0  &0  &\w^{20}&0  &0  &0  &\w^{20}&0  &0  &0  &\w^{20}\\
0  &0  &0  &-\w^{7}&0  &0  &0  &-\w^{15}&0  &0  &0  &\w^{23}&0  &0  &0  &-\w^{31}\\
0  &\w^{28}&0  &0  &0  &\w^{12}&0  &0  &0  &\w^{28}&0  &0  &0  &\w^{12}&0  &0  \\
0  &-\w^{23}&0  &0  &0  &-\w^{15}&0  &0  &0  &-\w^{7}&0  &0  &0  &-\w^{31}&0  &0  \\
0  &\w^{20}&0  &0  &0  &\w^{20}&0  &0  &0  &\w^{20}&0  &0  &0  &\w^{20}&0  &0  \\
0  &-\w^{27}&0  &0  &0  &-\w^{3}&0  &0  &0  &-\w^{11}&0  &0  &0  &-\w^{19}&0  &0  \\
\end{array}
\right).
\end{alignat*}
\end{footnotesize}
\caption{A change of basis for dimension $N=16$ resulting in a Zauner
  matrix which is a permutation, where $\tau=-\exp(\pi i/16)$. \label{fig:base_change16}}
\end{figure}

\begin{figure}[htp]
\begin{footnotesize}
\begin{alignat*}{5}\let\w\tau\arraycolsep0.5\arraycolsep
X=\left(\begin{array}{*{16}{c}}
0  &0  &0  &0  &0  &0  &0  &0  &0  &0  &0  &0  &\w^{4}&0  &0  &0  \\
0  &0  &0  &0  &0  &0  &0  &0  &0  &0  &0  &0  &0  &-\w^{9}&0  &0  \\
0  &0  &0  &0  &0  &0  &0  &0  &0  &0  &0  &0  &0  &0  &1&0  \\
0  &0  &0  &0  &0  &0  &0  &0  &0  &0  &0  &0  &0  &0  &0  &-\w^{5}\\
0  &0  &0  &0  &0  &0  &0  &0  &\w^{28}&0  &0  &0  &0  &0  &0  &0  \\
0  &0  &0  &0  &0  &0  &0  &0  &0  &-\w^{13}&0  &0  &0  &0  &0  &0  \\
0  &0  &0  &0  &0  &0  &0  &0  &0  &0  &\w^{20}&0  &0  &0  &0  &0  \\
0  &0  &0  &0  &0  &0  &0  &0  &0  &0  &0  &-\w^{25}&0  &0  &0  &0  \\
\w^{20}&0  &0  &0  &0  &0  &0  &0  &0  &0  &0  &0  &0  &0  &0  &0  \\
0  &-\w^{27}&0  &0  &0  &0  &0  &0  &0  &0  &0  &0  &0  &0  &0  &0  \\
0  &0  &1&0  &0  &0  &0  &0  &0  &0  &0  &0  &0  &0  &0  &0  \\
0  &0  &0  &-\w^{31}&0  &0  &0  &0  &0  &0  &0  &0  &0  &0  &0  &0  \\
0  &0  &0  &0  &\w^{28}&0  &0  &0  &0  &0  &0  &0  &0  &0  &0  &0  \\
0  &0  &0  &0  &0  &-\w^{23}&0  &0  &0  &0  &0  &0  &0  &0  &0  &0  \\
0  &0  &0  &0  &0  &0  &\w^{12}&0  &0  &0  &0  &0  &0  &0  &0  &0  \\
0  &0  &0  &0  &0  &0  &0  &-\w^{27}&0  &0  &0  &0  &0  &0  &0  &0 
\end{array}\right)
\end{alignat*}
\begin{alignat*}{5}\let\w\tau\arraycolsep0.5\arraycolsep
Z=\left(\begin{array}{*{16}{c}}
0  &1&0  &0  &0  &0  &0  &0  &0  &0  &0  &0  &0  &0  &0  &0  \\
0  &0  &\w^{12}&0  &0  &0  &0  &0  &0  &0  &0  &0  &0  &0  &0  &0  \\
0  &0  &0  &\w^{20}&0  &0  &0  &0  &0  &0  &0  &0  &0  &0  &0  &0  \\
1&0  &0  &0  &0  &0  &0  &0  &0  &0  &0  &0  &0  &0  &0  &0  \\
0  &0  &0  &0  &0  &\w^{4}&0  &0  &0  &0  &0  &0  &0  &0  &0  &0  \\
0  &0  &0  &0  &0  &0  &\w^{28}&0  &0  &0  &0  &0  &0  &0  &0  &0  \\
0  &0  &0  &0  &0  &0  &0  &\w^{12}&0  &0  &0  &0  &0  &0  &0  &0  \\
0  &0  &0  &0  &\w^{4}&0  &0  &0  &0  &0  &0  &0  &0  &0  &0  &0  \\
0  &0  &0  &0  &0  &0  &0  &0  &0  &-\w^{23}&0  &0  &0  &0  &0  &0  \\
0  &0  &0  &0  &0  &0  &0  &0  &0  &0  &-\w^{5}&0  &0  &0  &0  &0  \\
0  &0  &0  &0  &0  &0  &0  &0  &0  &0  &0  &-\w^{19}&0  &0  &0  &0  \\
0  &0  &0  &0  &0  &0  &0  &0  &-\w^{9}&0  &0  &0  &0  &0  &0  &0  \\
0  &0  &0  &0  &0  &0  &0  &0  &0  &0  &0  &0  &0  &-\w^{7}&0  &0  \\
0  &0  &0  &0  &0  &0  &0  &0  &0  &0  &0  &0  &0  &0  &-\w^{5}&0  \\
0  &0  &0  &0  &0  &0  &0  &0  &0  &0  &0  &0  &0  &0  &0  &-\w^{27}\\
0  &0  &0  &0  &0  &0  &0  &0  &0  &0  &0  &0  &-\w^{1}&0  &0  &0  \\
\end{array}\right)
\end{alignat*}
\end{footnotesize}
\caption{Generators $X$ and $Z$ of the Weyl-Heisenberg group in
  dimension $N=4^2$ in the adapted basis.  Here $\tau=-\exp(\pi
  i/16)$ denotes a primitive $32$nd root of
  unity.\label{fig:transformed_XZ}}
\end{figure}

In this basis, the Zauner matrix is a permutation matrix. Hence a
fiducial vector is of the form
\begin{align}
|\psi\rangle={}&
 x_0(|0\rangle+|2\rangle+|6\rangle)
+x_1(|1\rangle+|9\rangle+|10\rangle)
+x_3(|3\rangle+|14\rangle+|15\rangle)\nonumber\\
\label{eq:fiducial16} \\ 
&+x_4 |4\rangle
+x_5(|5\rangle+|11\rangle+|12\rangle)
+x_7(|7\rangle+|8\rangle+|13\rangle)\nonumber.
\end{align}
In order to solve the equations for these six complex variables $x_i$,
one of which can assumed to be real, we followed the approach
described in \cite{Grassl}. Computing a Gr\"obner basis modulo a
single 23-bit prime using Magma \cite{Magma} took about three days and
required about 30~GB of memory.  The polynomials in a Gr\"obner basis
with respect to so-called grevlex order have coefficients with 90
digits in the numerators and denominators.  Changing to lexicographic
order which is used to solve the equations, the coefficients grow to
some 900 digits. Nonetheless, we succeed to obtain a less complex
representation of a solution. The solutions are given in a number
field
\begin{equation}
{\mathbb{K}}={\mathbb{Q}}(\sqrt{2},\sqrt{13},\sqrt{17},r_2,r_3,t_1,t_2,t_3,t_4,\sqrt{-1}),
\end{equation}
of degree 1024, where
\begin{align}
r_2={}&\sqrt{\sqrt{221}-11},\qquad\qquad 
r_3=\sqrt{15+\sqrt{17}},\displaybreak[0] \nonumber\\
t_1={}&\sqrt{15+(4-\sqrt{17})r_3 -3\sqrt{17}},\displaybreak[0] \\
t_2^2={}&(((3-5\sqrt{17})\sqrt{13}+(39\sqrt{17}-65))r_3+((16\sqrt{17}-72)\sqrt{13}+936))t_1 \nonumber\\
&-208\sqrt{13}+2288,\displaybreak[0] \nonumber\\
t_3={}&\sqrt{2-\sqrt{2}},\qquad\qquad
t_4=\sqrt{2+t_3}. \nonumber
\end{align}
Note that we do not explicitly use a $32$nd root of unity, it can be
expressed as
\begin{equation}
\omega_{32}=\frac{1}{2}((\sqrt{2}(1-t_3)-1)t_4-t_4\sqrt{-1}).
\end{equation}
The Galois group of ${\mathbb{K}}$ is isomorphic to
$C_8\times((C_2\times C_2 \times C_{16})\rtimes C_2)$, and
$\mathbb{K}$ is an Abelian extension of ${\mathbb{Q}}(\sqrt{221})$.
The coefficients of a non-normalized fiducial vector of
eq. (\ref{eq:fiducial16}) are as follows:
\begin{footnotesize}
\begin{align}
x_0={}&-\frac{40}{13}\sqrt{13}r_3t_1t_2\\
x_1={}&\Bigl(\bigl((21\sqrt{2}+22\sqrt{13}+16\sqrt{17}+5\sqrt{26}+5\sqrt{34}+4\sqrt{221}+\sqrt{442}+74)r_2r_3\displaybreak[0] \nonumber \\
      &+(-77\sqrt{2}-26\sqrt{13}-18\sqrt{17}-33\sqrt{26}-19\sqrt{34}+2\sqrt{221}-7\sqrt{442}+42)r_2\displaybreak[0] \nonumber\\
      &+(-45\sqrt{2}+30\sqrt{13}-10\sqrt{17}+15\sqrt{26}+5\sqrt{34}+10\sqrt{221}+5\sqrt{442}-30)r_3\displaybreak[0] \nonumber\\
      &+(30\sqrt{13}+30\sqrt{17}+10\sqrt{221}-70)\bigr)t_1t_4\displaybreak[0] \nonumber\\
      &+\bigl((3\sqrt{2}+3\sqrt{13}+9\sqrt{17}-\sqrt{26}+7\sqrt{34}+11\sqrt{221}+3\sqrt{442}+121)r_2r_3\displaybreak[0]\\
      &+(-82\sqrt{2}-88\sqrt{13}-24\sqrt{17}-74\sqrt{26}-2\sqrt{34}+24\sqrt{221}-2\sqrt{442}+264)r_2\displaybreak[0] \nonumber\\
      &+(175\sqrt{2}+80\sqrt{13}-20\sqrt{17}+75\sqrt{26}-25\sqrt{34}-5\sqrt{442}+380)r_3\displaybreak[0] \nonumber\\
      &+(200\sqrt{2}+220\sqrt{13}-300\sqrt{17}+160\sqrt{26}-160\sqrt{34}-20\sqrt{221}-40\sqrt{442}+180)\bigr)t_4\Bigr)\sqrt{-1}\displaybreak[2] \nonumber\\
      &+\bigl((-10\sqrt{2}-15\sqrt{13}-15\sqrt{17}-6\sqrt{26}-8\sqrt{34}-\sqrt{221}-21)r_2r_3\displaybreak[0] \nonumber\\
      &+(55\sqrt{2}+16\sqrt{13}+28\sqrt{17}+7\sqrt{26}+21\sqrt{34}+8\sqrt{221}+5\sqrt{442}+108)r_2\displaybreak[0] \nonumber\\
      &+(70\sqrt{2}+10\sqrt{26}+80)r_3\displaybreak[0] \nonumber\\
      &-10\sqrt{2}-130\sqrt{13}-250\sqrt{17}-70\sqrt{26}-70\sqrt{34}-30\sqrt{221}-10\sqrt{442}-630\bigr)t_1t_4\displaybreak[0] \nonumber\\
      &+\bigl((10\sqrt{2}-51\sqrt{13}-33\sqrt{17}-24\sqrt{26}-22\sqrt{34}+\sqrt{221}-29)r_2r_3\displaybreak[0] \nonumber\\
      &+(320\sqrt{2}-4\sqrt{13}+28\sqrt{17}+8\sqrt{26}+4\sqrt{34}+44\sqrt{221}+20\sqrt{442}+524)r_2\displaybreak[0] \nonumber\\
      &+(265\sqrt{2}-30\sqrt{13}-50\sqrt{17}-15\sqrt{26}-35\sqrt{34}+10\sqrt{221}+5\sqrt{442}+310)r_3\displaybreak[0] \nonumber\\
      &+(260\sqrt{2}-200\sqrt{13}-560\sqrt{17}-100\sqrt{26}-260\sqrt{34}+20\sqrt{442}-600)\bigr)t_4\displaybreak[3] \nonumber\\
x_3={}&\Bigl(\bigl((-21\sqrt{2}-22\sqrt{13}-16\sqrt{17}-5\sqrt{26}-5\sqrt{34}-4\sqrt{221}-\sqrt{442}-74)r_2r_3\displaybreak[0] \nonumber\\
      &+(77\sqrt{2}+26\sqrt{13}+18\sqrt{17}+33\sqrt{26}+19\sqrt{34}-2\sqrt{221}+7\sqrt{442}-42)r_2\displaybreak[0] \nonumber\\
      &+(-45\sqrt{2}+30\sqrt{13}-10\sqrt{17}+15\sqrt{26}+5\sqrt{34}+10\sqrt{221}+5\sqrt{442}-30)r_3\displaybreak[0] \nonumber\\
      &+(30\sqrt{13}+30\sqrt{17}+10\sqrt{221}-70)\bigr)t_1t_4\displaybreak[0] \nonumber\\
      &+\bigl((-3\sqrt{2}-3\sqrt{13}-9\sqrt{17}+\sqrt{26}-7\sqrt{34}-11\sqrt{221}-3\sqrt{442}-121)r_2r_3\displaybreak[0] \nonumber \\
      &+(82\sqrt{2}+88\sqrt{13}+24\sqrt{17}+74\sqrt{26}+2\sqrt{34}-24\sqrt{221}+2\sqrt{442}-264)r_2\displaybreak[0] \nonumber \\
      &+(175\sqrt{2}+80\sqrt{13}-20\sqrt{17}+75\sqrt{26}-25\sqrt{34}-5\sqrt{442}+380)r_3\displaybreak[0] \nonumber \\
      &+(200\sqrt{2}+220\sqrt{13}-300\sqrt{17}+160\sqrt{26}-160\sqrt{34}-20\sqrt{221}-40\sqrt{442}+180)\bigr)t_4\Bigr)\sqrt{-1}\displaybreak[2] \nonumber \\
      &+\bigl((10\sqrt{2}+15\sqrt{13}+15\sqrt{17}+6\sqrt{26}+8\sqrt{34}+\sqrt{221}+21)r_2r_3\displaybreak[0]\\
      &+(-55\sqrt{2}-16\sqrt{13}-28\sqrt{17}-7\sqrt{26}-21\sqrt{34}-8\sqrt{221}-5\sqrt{442}-108)r_2\displaybreak[0] \nonumber \\
      &+(70\sqrt{2}+10\sqrt{26}+80)r_3\displaybreak[0] \nonumber \\
      &-10\sqrt{2}-130\sqrt{13}-250\sqrt{17}-70\sqrt{26}-70\sqrt{34}-30\sqrt{221}-10\sqrt{442}-630\bigr)t_1t_4\displaybreak[0] \nonumber \\
      &+\bigl((-10\sqrt{2}+51\sqrt{13}+33\sqrt{17}+24\sqrt{26}+22\sqrt{34}-\sqrt{221}+29)r_2r_3\displaybreak[0] \nonumber \\
      &+(-320\sqrt{2}+4\sqrt{13}-28\sqrt{17}-8\sqrt{26}-4\sqrt{34}-44\sqrt{221}-20\sqrt{442}-524)r_2\displaybreak[0] \nonumber \\
      &+(265\sqrt{2}-30\sqrt{13}-50\sqrt{17}-15\sqrt{26}-35\sqrt{34}+10\sqrt{221}+5\sqrt{442}+310)r_3\displaybreak[0] \nonumber \\
      &+(260\sqrt{2}-200\sqrt{13}-560\sqrt{17}-100\sqrt{26}-260\sqrt{34}+20\sqrt{442}-600)\bigr)t_4\displaybreak[3] \nonumber \\
x_4={}&\bigl((-\frac{11}{26}\sqrt{13}-\frac{1}{2}\sqrt{17}-\frac{3}{26}\sqrt{221}-\frac{1}{2})r_2r_3\displaybreak[3] \nonumber \\
      &+(10\sqrt{2}+\frac{20}{13}\sqrt{26}+\frac{10}{13}\sqrt{442})r_2\bigr)t_1t_2\sqrt{-1}\displaybreak[2]\\
      &+\bigl((\frac{11}{26}\sqrt{13}+\frac{1}{2}\sqrt{17}+\frac{3}{26}\sqrt{221}+\frac{1}{2})r_2r_3\displaybreak[0] \nonumber \\
      &+(10\sqrt{2}+\frac{20}{13}\sqrt{26}+\frac{10}{13}\sqrt{442})r_2\bigr)t_1t_2\displaybreak[3] \nonumber \\
x_5={}&\Bigl(\bigl((-\frac{75}{2}\sqrt{2}-4\sqrt{13}-2\sqrt{17}-\frac{25}{2}\sqrt{26}-\frac{15}{2}\sqrt{34}-\frac{5}{2}\sqrt{442}+10)r_2r_3\displaybreak[0]  \nonumber \\
      &+(-22\sqrt{2}-24\sqrt{13}-12\sqrt{17}+14\sqrt{26}+2\sqrt{34}-4\sqrt{221}-2\sqrt{442}-24)r_2\displaybreak[0] \nonumber \\
      &+(15\sqrt{2}-5\sqrt{13}-5\sqrt{17}+25\sqrt{26}-5\sqrt{34}-5\sqrt{221}+5\sqrt{442}+35)r_3\displaybreak[0] \nonumber \\
      &+(270\sqrt{2}+60\sqrt{13}+180\sqrt{17}+10\sqrt{26}+70\sqrt{34}+20\sqrt{221}+10\sqrt{442}+620)\bigr)t_1\displaybreak[0] \nonumber \\
      &+\bigl((-85\sqrt{2}-28\sqrt{13}-4\sqrt{17}-3\sqrt{26}+\sqrt{34}+4\sqrt{221}-5\sqrt{442}-36)r_2r_3\displaybreak[0] \nonumber \\
      &+(-190\sqrt{2}-86\sqrt{13}+22\sqrt{17}+34\sqrt{26}+22\sqrt{34}+22\sqrt{221}-10\sqrt{442}+122)r_2\displaybreak[0] \nonumber \\
      &+(300\sqrt{2}-60\sqrt{13}+40\sqrt{17}+40\sqrt{26}-20\sqrt{34}-220)r_3\displaybreak[0] \nonumber \\
      &+(650\sqrt{2}-60\sqrt{13}+460\sqrt{17}+110\sqrt{26}-130\sqrt{34}+60\sqrt{221}+10\sqrt{442}+660)\bigr)\Bigr)\sqrt{-1}\displaybreak[2] \nonumber \\
      &+\bigl((1\frac{1}{2}\sqrt{2}+23\sqrt{13}+19\sqrt{17}-\frac{3}{2}\sqrt{26}-\frac{9}{2}\sqrt{34}+3\sqrt{221}+\frac{1}{2}\sqrt{442}+63)r_2r_3\displaybreak[0]\\
      &+(152\sqrt{2}-20\sqrt{17}+28\sqrt{26}+24\sqrt{34}+4\sqrt{221}+12\sqrt{442}+64)r_2\displaybreak[0] \nonumber \\
      &+(-70\sqrt{2}-5\sqrt{13}+15\sqrt{17}+10\sqrt{34}-5\sqrt{221}+55)r_3\displaybreak[0] \nonumber \\
      &+(350\sqrt{2}-100\sqrt{13}-100\sqrt{17}+50\sqrt{26}+110\sqrt{34}-20\sqrt{221}+10\sqrt{442}+60)\bigr)t_1\displaybreak[0] \nonumber \\
      &+(43\sqrt{2}+28\sqrt{13}+24\sqrt{17}-23\sqrt{26}-19\sqrt{34}+8\sqrt{221}+3\sqrt{442}+108)r_2r_3\displaybreak[0] \nonumber \\
      &+(476\sqrt{2}-22\sqrt{13}-26\sqrt{17}+28\sqrt{26}+4\sqrt{34}+6\sqrt{221}+36\sqrt{442}+26)r_2\displaybreak[0] \nonumber \\
      &+(-170\sqrt{2}-20\sqrt{13}-40\sqrt{17}+50\sqrt{26}+10\sqrt{34}-10\sqrt{442}+60)r_3\displaybreak[0] \nonumber \\
      &+410\sqrt{2}-160\sqrt{13}-120\sqrt{17}+150\sqrt{26}+270\sqrt{34}-30\sqrt{442}+280\displaybreak[3] \nonumber \\
x_7={}&\Bigl(\bigl((-1\frac{1}{2}\sqrt{2}-23\sqrt{13}-19\sqrt{17}+\frac{3}{2}\sqrt{26}+\frac{9}{2}\sqrt{34}-3\sqrt{221}-\frac{1}{2}\sqrt{442}-63)r_2r_3\displaybreak[0]  \nonumber \\
      &+(-152\sqrt{2}+20\sqrt{17}-28\sqrt{26}-24\sqrt{34}-4\sqrt{221}-12\sqrt{442}-64)r_2\displaybreak[0] \nonumber \\
      &+(-70\sqrt{2}-5\sqrt{13}+15\sqrt{17}+10\sqrt{34}-5\sqrt{221}+55)r_3\displaybreak[0] \nonumber \\
      &+(350\sqrt{2}-100\sqrt{13}-100\sqrt{17}+50\sqrt{26}+110\sqrt{34}-20\sqrt{221}+10\sqrt{442}+60)\bigr)t_1\displaybreak[0] \nonumber \\
      &+\bigl((-43\sqrt{2}-28\sqrt{13}-24\sqrt{17}+23\sqrt{26}+19\sqrt{34}-8\sqrt{221}-3\sqrt{442}-108)r_2r_3\displaybreak[0] \nonumber \\
      &+(-476\sqrt{2}+22\sqrt{13}+26\sqrt{17}-28\sqrt{26}-4\sqrt{34}-6\sqrt{221}-36\sqrt{442}-26)r_2\displaybreak[0]\\
      &+(-170\sqrt{2}-20\sqrt{13}-40\sqrt{17}+50\sqrt{26}+10\sqrt{34}-10\sqrt{442}+60)r_3\displaybreak[0] \nonumber \\
      &+(410\sqrt{2}-160\sqrt{13}-120\sqrt{17}+150\sqrt{26}+270\sqrt{34}-30\sqrt{442}+280)\bigr)\Bigr)\sqrt{-1}\displaybreak[2] \nonumber \\
      &+\bigl((-\frac{75}{2}\sqrt{2}-4\sqrt{13}-2\sqrt{17}-\frac{25}{2}\sqrt{26}-\frac{15}{2}\sqrt{34}-\frac{5}{2}\sqrt{442}+10)r_2r_3\displaybreak[0] \nonumber \\
      &+(-22\sqrt{2}-24\sqrt{13}-12\sqrt{17}+14\sqrt{26}+2\sqrt{34}-4\sqrt{221}-2\sqrt{442}-24)r_2\displaybreak[0] \nonumber \\
      &+(-15\sqrt{2}+5\sqrt{13}+5\sqrt{17}-25\sqrt{26}+5\sqrt{34}+5\sqrt{221}-5\sqrt{442}-35)r_3\displaybreak[0] \nonumber \\
      &-270\sqrt{2}-60\sqrt{13}-180\sqrt{17}-10\sqrt{26}-70\sqrt{34}-20\sqrt{221}-10\sqrt{442}-620\bigr)t_1\displaybreak[0] \nonumber \\
      &+(-85\sqrt{2}-28\sqrt{13}-4\sqrt{17}-3\sqrt{26}+\sqrt{34}+4\sqrt{221}-5\sqrt{442}-36)r_2r_3\displaybreak[0] \nonumber \\
      &+(-190\sqrt{2}-86\sqrt{13}+22\sqrt{17}+34\sqrt{26}+22\sqrt{34}+22\sqrt{221}-10\sqrt{442}+122)r_2\displaybreak[0] \nonumber \\
      &+(-300\sqrt{2}+60\sqrt{13}-40\sqrt{17}-40\sqrt{26}+20\sqrt{34}+220)r_3\displaybreak[0] \nonumber \\
      &-650\sqrt{2}+60\sqrt{13}-460\sqrt{17}-110\sqrt{26}+130\sqrt{34}-60\sqrt{221}-10\sqrt{442}-660 \nonumber 
\end{align}
\end{footnotesize}%
Fiducial vectors for SICs with respect to the standard representation
of the Weyl-Heisenberg group are given in Appendix~C.  It turns out
that the two numerical solutions with orbits labelled $16a$ and $16b$ in \cite{Scott} are
related by the Galois automorphism of $\mathbb{K}$ induced by
simultaneously changing the signs of $\sqrt{13}$ and $\sqrt{17}$.

\vspace{10mm}

{\bf 8. A remark on Mutually Unbiased Bases}

\vspace{5mm}

\noindent Although our main emphasis has been on SICs, we note that square dimensions are special also in the Mutually Unbiased Bases (MUB) problem. In dimensions $N=n^2$, Wocjan and Beth have shown that one can construct sets of Mutually Unbiased Bases using $n$ by $n$ Latin squares \cite{Wocjan}. In this section, we examine a variant of the MUB problem making use of the phase-permutation basis.

 
We would like to find all vectors, and all bases, unbiased with 
respect to the two eigenbases defined by two complementary cyclic subgroups 
of the Weyl-Heisenberg group. This problem has been studied in connection with the MUB 
existence problem in dimension six \cite{Markus} and in another guise is also known as the cyclic $N$-roots problem \cite{Bjorck}. The two bases 
can be taken to be the standard basis $|u\rangle_0$, the eigenbasis of 
the subgroup generated by $Z$, and the 
Fourier basis $|u\rangle_\infty$, which is the eigenbasis of the subgroup 
generated by $X$. (The labels $0$ and $\infty$ 
do have a logical explanation \cite{Chris}.) We are looking for 
all vectors $|\psi\rangle$ such that 

\begin{equation}\label{cyclic N} |\langle \psi|u\rangle_0|^2 = |\langle \psi|u\rangle_\infty 
|^2 = \frac{1}{N} \ , \end{equation}

\noindent for all $N$ values of $u$. The answer is known for $N \leq 9$, 
and it is also known that such vectors always belong to a complete basis 
unbiased with respect to the two bases we start out with. 
Here we just want to report what this problem looks like when $N = n^2$ 
and the phase-permutation basis is used.  

The two eigenbases are now given by 
\begin{equation} |a+nb\rangle_0 = \frac{1}{\sqrt{n}} 
\sum_{r=0}^{n-1}\sigma^{-br}|r,a\rangle 
\ , \hspace{10mm} Z|a+nb\rangle_0 =  \omega^{a+nb}|a+nb\rangle_0 \ \end{equation} 
\begin{equation} |a+nb\rangle_\infty = \frac{1}{\sqrt{n}} \sum_{r=0}^{n-1}\sigma^{-br}
\omega^{-ar}|a,r\rangle \ , \hspace{6mm} X|a+nb\rangle_\infty =  \omega^{a+nb}|a+nb\rangle_\infty \ . \end{equation}
\noindent We look for bases unbiased with respect to this pair, of the form 

\begin{equation} |a+nb\rangle_k = \frac{1}{\sqrt{n}}\sum_{r=0}^{n-1} 
\omega_k(r,a,b)|r, \lambda_k(r,a)\rangle \ . \end{equation}

\noindent Here $\omega_k(r,a,b)$ is a phase factor, and $\lambda_k$ is
a map from $\ZZ_{n}\times \ZZ_{n}$ to $\ZZ_{n}$, where $\ZZ_{n}$ is
the set of integers modulo $n$.  


Unbiasedness with respect to the standard basis holds if and only if the function
$\lambda_k(r,a)$ is injective for fixed $a$. To see this, we argue by contradiction. Suppose $\lambda _{k}(r,a)$ is not injective for some fixed $a$, then there exists an
integer $x\in {\mathbb{Z}}_{n}$ such that $\lambda _{k}(r,a)\neq x$ for all $r$. Now consider the inner product%
\begin{eqnarray}
\langle x+nb^{\prime }|_{0}\;a+nb\rangle _{k} &=&\frac{1}{n}%
\sum_{r=0}^{n-1}q^{b^{\prime }r}\omega _{k}(r,a,b)\delta _{z,\lambda
_{k}(r,a)} \\
&=&0 \,,
\end{eqnarray}%
since by assumption, $\delta _{x,\lambda _{k}(r,a)}=0$ for all $r.$ Hence,
if $\lambda _{k}$ is not injective there is at least one vector from the
standard basis that is orthogonal to the new basis. There are is no way to
choose the phases $\omega _{k}$ to make it MU to every vector in the
standard basis.

Similarly, unbiasedness with respect to the Fourier basis holds if and only if the function is
injective for fixed $r$. This means that the function $\lambda_k(r,a)$
defines a Latin square, an $n$ by $n$ array such that each row and
column contain the symbols from an $n$-letter alphabet exactly once.

As an example, let $n = p$ be a prime number. Then the Weyl-Heisenberg group 
contains $p+1$ cyclic subgroups altogether. Two of them were accounted for 
from the start, and the remaining $p-1$ examples give rise to the choice 

\begin{equation} \lambda_k(r,a) = a + kr \ , \hspace{8mm} k \in \{ 1, 2, \dots, 
p-1\} \ . \end{equation}

\noindent By inspection one finds that these $p-1$ bases are maximally 
entangled. In addition, the Latin squares that define the bases are Mutually Orthogonal, which ensures that we have a collection of 
$p+1$ Mutually Unbiased Bases.     

The fact that $x-1$ Mutually Orthogonal Latin squares of order $n$ give 
rise to $x+1$ Mutually Unbiased Bases---regardless of whether they originate 
from group theory or not---was first observed by Wocjan and Beth \cite{Wocjan}. 
The only new observation here is that Latin squares appear naturally in the cyclic $N$-roots problem when the phase-permutation basis is used.

To illustrate the idea, consider dimension $N=2^2$. The two eigenbases are given by 
\begin{eqnarray}
|0\rangle _{0}=\frac{1}{\sqrt{2}}\left( |0,0\rangle +|1,0\rangle \right) \;\;\;\;
&|2\rangle _{0}=\frac{1}{\sqrt{2}}\left( |0,0\rangle -|1,0\rangle \right)  
 \\
|1\rangle _{0}=\frac{1}{\sqrt{2}}\left( |0,1\rangle +|1,1\rangle \right) \;\;\;\;
&|3\rangle _{0}=\frac{1}{\sqrt{2}}\left( |0,1\rangle -|1,1\rangle \right) ,
\label{N=4 basis 1}
\end{eqnarray}%
and 
\begin{eqnarray}
|0\rangle _{\infty }=\frac{1}{\sqrt{2}}\left( |0,0\rangle +|0,1\rangle
\right) \;\;\;\; &|2\rangle _{\infty }=\frac{1}{\sqrt{2}}\left( |0,0\rangle
-|0,1\rangle \right)    \\
|1\rangle _{\infty }=\frac{1}{\sqrt{2}}\left( |1,0\rangle -i|1,1\rangle
\right) \;\;\;\; &|3\rangle _{\infty }=\frac{1}{\sqrt{2}}\left( |1,0\rangle
+i|1,1\rangle \right) .  \label{N=4 basis 2}
\end{eqnarray}%
The Latin square $\lambda _1=a+r$, then generates the third basis
\begin{eqnarray}
|0\rangle _{1}=\frac{1}{\sqrt{2}}\left( |0,0\rangle +\theta _{0}|1,1\rangle
\right) \;\;\;\; &|2\rangle _{1}=\frac{1}{\sqrt{2}}\left( |0,0\rangle +\theta
_{2}|1,1\rangle \right)  \\
|1\rangle _{1}=\frac{1}{\sqrt{2}}\left( |0,1\rangle +\theta _{1}|1,0\rangle
\right) \;\;\;\; &|3\rangle _{1}=\frac{1}{\sqrt{2}}\left( |0,1\rangle +\theta
_{3}|1,0\rangle \right) ,
\end{eqnarray}%
unbiased with respect to eqs. (\ref{N=4 basis 1}) and (\ref{N=4 basis 2}).
We have removed an overall phase from each vector leaving the remaining free
phases $\theta _{0},\ldots ,\theta _{3}.$ The conditions for the vectors to
form a basis are simply that $1+\theta _{0}\overline{\theta _{2}}=0$ and $%
1+\theta _{1}\overline{\theta _{3}}=0.$

In dimension $N=4$, this method constructs the complete set of solutions to
the cyclic $N$-roots problem. In dimension 9 however, whilst we find the two
parameter family of solutions, there are 6,156 other isolated points \cite{Faugere}.


\vspace{10mm}
\newpage
{\bf 9. Conclusions and questions}

\vspace{5mm}

\noindent Our main result is that the entire Clifford group admits a 
representation using only monomial phase-permutation matrices if and only 
if the dimension is a square number. We also gave such a representation 
explicitly. 

In the course of the proof we proved some theorems about the Weyl-Heisenberg 
and Clifford groups that we suspect are known, but which we were unable to 
find in the literature.

We used this representation to gain some insight into the SIC problem. 
It shares with the standard representation the elegant property that 
a SIC when orthogonally projected onto the simplex spanned by the basis 
within the body of density matrices is again a regular simplex, with 
$N$ rather than $N^2$ vertices. 
Expressions for the SIC fiducials 
are much simplified,  but are still considerably hard to calculate when the 
dimension exceeds $2^2$. At least the case $N=4$ is now trivial, and we can now solve the case $N=9$ by hand. More significantly, we find for the first time an exact solution to the SIC problem in dimension $N=16$.

We have applied the phase-permutation representation to the problem of finding all vectors unbiased with respect to both the standard and the Fourier bases. Families of solutions can then be constructed naturally from Latin squares. Finally, we note that quantum mechanics in square-dimensional Hilbert spaces is of particular importance because they admit bipartite entanglement; we therefore expect that the phase-permutation representation will have many other applications.   

\

\

\noindent \underline{Acknowledgments}: We thank \AA sa Ericsson for
the idea of Section 8 and Steve Donkin for discussions on Appendix B.
The authors gratefully acknowledge the hospitality of the Nordita
workshop on the Foundations of Quantum Mechanics. We thank Berge Englert for inviting IB to CQT, which led to some motivating discussions.

DMA was supported in part by the U.~S. Office of Naval Research (Grant
No.\ N00014-09-1-0247). Research at Perimeter Institute is supported
by the Government of Canada through Industry Canada and by the
Province of Ontario through the Ministry of Research \& Innovation.
IB is supported by the Swedish Research Council under contract VR
621-2007-4060.  SB was supported by the EU FP7 FET-Open research
project COMPAS (Contract No. 212008).  DG gratefully acknowledges
support by the Institut Mittag-Leffler (Djursholm, Sweden), where his
contribution to this work was done. DG's research is supported by the
German Science Foundation (DFG grants CH 843/1-1 and CH 843/2-1) and
the Swiss National Science Foundation.


\vspace{10mm}

{\bf Appendix A: Action of the Clifford group}

\vspace{5mm} 

\noindent {\bf Lemma 3}: {\sl
	The action of the Clifford group on $H(N)/Z(N)\simeq \ZZ_N^2$ is
	isomorphic to $SL(2,N)$.
}

\ 

\noindent {\sl Proof}: 
Let $G:\ZZ_N^2\to\ZZ_N^2$ be a transformation induced by the action of
a Clifford unitary on $H(N)/Z(N)$. First, we show that $G$ must be an
element of $SL(2,N)$. 

That $G$ is linear follows from the fact that $H(N)/Z(N)$ is a
projective representation of $\ZZ_N^2$. Now consider the following
commutation relation, which is a simple consequence of (\ref{symp}):
\begin{equation}\label{eqn:commutation}
	D_{ij} D_{kl} = \omega^{kj - il} D_{kl} D_{ij}
\end{equation}
for $i,j,k,l\in\ZZ_N$. Conjugate every matrix appearing in the
relation above by the Clifford unitary $U_G$. All the phase factors
$\tau^{k'}$ appearing in the definition (\ref{defcliff}) cancel,
because they occur on both sides of the equality. With
$(i',j')=G(i,j)$ and $(k',l')=G(k,l)$, we conclude that
\begin{equation}
	\omega^{k j - i l} = \omega^{k'j' - i'l'}.
\end{equation}
Because $\omega$ has order $N$, $G$ preserves symplectic inner
products modulo $N$. Thus, $G\in SL(2,N)$ as claimed.

Next, we have to show that every transformation in $SL(2,N)$ can be
realized. For $N$ odd, this is the content of (\ref{eqn:UG}) proven in
Ref.~\cite{Marcus}. Hence, we only need to consider the case of even
$N$. In this case, (\ref{eqn:UG}) says that if $G\in SL(2,\bar N)$,
then $G\,\mbox{mod}\,N$ may be realized as a transformation of
$H(N)/Z(N)$. Therefore, what remains to be shown is that every matrix
$G$ in $SL(2,N)$ can be written as $\bar G\,\mbox{mod}\,N$ for some
$\bar G \in SL(2,2 N)$.

Write $N$ as $N=2^l n$ for $n$ odd. In Appendix B, we show
that
\begin{equation}
	H(N) \simeq H(2^l) \times H(n), \qquad
	SL(2,\bar N) \simeq SL(2, 2 2^{l}) \times SL(2,n).
\end{equation}
What is more, ``the even and the odd parts do not mix'' in the sense
that $SL(2,2 2^{l})$ only acts on $H(2^l)$ and $SL(2,n)$ only acts on
$H(n)$. Therefore, we need to prove the claim only for the case $N=2^l$.

So let $G\in SL(2,N)$. Then $\det G = k N + 1$ for some integer $k$.
If $k$ is even, then $\det G \equiv 1 \mbox{ mod }2 N$ and
therefore $G\in SL(2,2N)$, so we are done. Thus we assume that $k$
is odd. Not all matrix elements of $G$ are even, for
then the range of $G$ would consist only of vectors with even
components. This would contradict that fact that $G$ is invertible.
Assume for now that $\alpha$, the top left matrix element of $G$, is
odd (we label the matrix elements of $G$ as in (\ref{eqn:Gmatrix})).
Then it has an inverse $\alpha^{-1}$ modulo $2N$. Now let
\begin{equation}
	\bar G =
	\left(
		\begin{array}{cc}
			\alpha & \beta \\
			\gamma & \delta + \alpha^{-1} N
		\end{array}
	\right).
\end{equation}
Then 
\begin{equation}
	\det \bar G =  \det G + \alpha \alpha^{-1}N 
	= (k+\alpha \alpha^{-1}) N +1 \equiv 1 \quad\mbox{mod}\,2 N.
\end{equation}
Thus $\bar G\in
SL(2,2N)$. The cases where one of the other matrix elements of
$G$ is odd are treated analogously.  $\Box$

\vspace{10mm}

{\bf Appendix B: Tensor Product Representation}

\vspace{5mm}


\noindent The Weyl-Heisenberg group in dimension $N$ is defined in terms of three
generators $\tilde{X},$ $\tilde{Z}$ and $\tilde{\tau}$ and the relations
between them. We denote the abstract group elements with a tilde; their
standard unitary representations, defined in eq. (\ref{Steve's7}), appear without. In odd
dimensions, the Weyl-Heisenberg group, $H(N)$ is given by 
\begin{eqnarray}
\left\langle \tilde{X},\tilde{Z},\tilde{\tau}\; : \; \tilde{X}^{N}=\tilde{Z%
}^{N}=\tilde{\tau}^{N}=1, 
\;\tilde{Z}\tilde{X}=\tilde{\tau}\tilde{X}\tilde{Z}%
,\;\tilde{\tau}\tilde{X}=\tilde{X}\tilde{\tau},\;\tilde{\tau}\tilde{Z}=%
\tilde{Z}\tilde{\tau} \right\rangle, 
\end{eqnarray}
whilst for $N$ even, we choose to enlarge the centre and define $H(N)$ to
be 
\begin{equation}
\left\langle \tilde{X},\tilde{Z},\tilde{\tau}\;:\;\tilde{X}^{N}=\tilde{Z%
}^{N}=\tilde{\tau}^{2N}=1,\;\tilde{Z}\tilde{X}=\tilde{\tau}^{2}\tilde{X}%
\tilde{Z},\;\tilde{\tau}\tilde{X}=\tilde{X}\tilde{\tau},\;\tilde{\tau}\tilde{%
Z}=\tilde{Z}\tilde{\tau}\right\rangle . 
\end{equation}

When the dimension has the prime factorization $N=n_{1}n_{2}\ldots n_{r},$
where $n_{j}=p_{j}^{u_{j}}$, the group is a direct product of smaller
groups, 
\begin{equation}
H(N)=H(n_{1})\times H(n_{2})\times \cdots \times H(n_{r}). 
\end{equation}
To see this, we construct an isomorphism as follows. Let the group elements
of $H(n_{j})$ be generated by $\tilde{X}_{j},$ $\tilde{Z}_{j}$ and $\tilde{%
\tau}_{j}$ and define elements of $H(n_{1})\times \cdots \times H(n_{r})$ as%
\begin{eqnarray}
x &=&(\tilde{X}_{1},\ldots ,\tilde{X}_{r}), \nonumber \\
z &=&(\tilde{Z}_{1},\ldots ,\tilde{Z}_{r}), \\
t &=&(\tilde{\tau}_{1},\ldots ,\tilde{\tau}_{r}). \nonumber
\end{eqnarray}%
The elements $x,$ $z$ and $t$ satisfy the relations for the group $H(N)$
since, for example, 
\begin{equation}
x^{N}=((\tilde{X}_{1}^{n_{1}})^{N/n_{1}},(\tilde{X}_{2}^{n_{2}})^{N/n_{2}},%
\ldots ,(\tilde{X}_{2}^{n_{r}})^{N/n_{r}})=(1,1,\ldots ,1). 
\end{equation}
Therefore, the map 
\begin{equation}
\theta :\tilde{X}^{a}\tilde{Z}^{b}\tilde{\tau}^{c}\rightarrow
x^{a}z^{b}t^{c} 
\end{equation}
is a homomorphism.

The image of $\theta $ is given by all elements of the form $x^{a}z^{b}t^{c}$
and we now show that it contains the group $H(n_{1})\times \cdots \times
H(n_{r}).$ The Chinese remainder theorem tells us that since $n_{j}$ and $%
n_{k}$ are coprime for all $j\neq k$ there exists an integer $\lambda _{1}$
such that $\lambda _{1}\equiv 1 \textrm{ mod } n_{1}$ and $\lambda _{1}\equiv 0%
\textrm{ mod } n_{j}$ for $j=2,\ldots ,r$. The integer $\lambda _{1}$ picks out
the first component of $x,$ 
\begin{equation}
x^{\lambda _{1}}=(\tilde{X}_{1}^{\lambda _{1}},\tilde{X}_{2}^{\lambda
_{1}},\ldots ,\tilde{X}_{r}^{\lambda _{1}})=(\tilde{X}_{1},1,\ldots ,1).
\end{equation}
In the same way, there exist integers, $\lambda _{2},\ldots ,\lambda _{r}$
and $\mu _{1},\ldots ,\mu _{r}$ such that 
\begin{eqnarray}
x^{\lambda _{j}} &=&(1,\ldots ,1,\tilde{X}_{j},1,\ldots ,1) \nonumber \\
z^{\mu _{j}} &=&(1,\ldots ,1,\tilde{Z}_{j},1,\ldots ,1).
\end{eqnarray}

The components of the element $t$ are computed modulo $\bar{n}_{j}$ so we
need to modify our argument slightly. In even dimensions, the Chinese
remainder theorem still applies since only one of the factors, say $n_{1}$,
is even. The integer $2n_{1}$ is therefore coprime to $n_{j}$ for all $j,$
and we can again find integers $\nu _{1},\ldots ,\nu _{r}$ such that 
\begin{equation}
t^{\nu _{j}}=(1,\ldots ,1,\tilde{\tau}_{j},1,\ldots ,1).
\end{equation}
Finally, the size of the two groups are equal, $|H(N)|=N^{2}\bar{N}%
=|H(n_{1})\times \cdots \times H(n_{r})|$ so $\theta $ is an isomorphism.

Now for the Clifford group $C(N)$. We use the fact that $C(N)$ is the
semi-direct product of $H(N)$ and $SL(2,\bar{N})$. We start by
taking elements of $SL(2,\bar{N})$ and computing their
components modulo $\bar{n}_{j},$ that is,%
\begin{equation}
F_{j}\equiv \left( 
\begin{array}{cc}
\alpha _{j} & \beta _{j} \\ 
\gamma _{j} & \delta _{j}%
\end{array}%
\right) ,
\end{equation}
where $\alpha _{j}=\alpha \textrm{ mod }\bar{n}_{j},$ $\beta _{j}=\beta \textrm{ mod }\bar{n}_{j},$ $\gamma _{j}=\gamma \textrm{ mod }\bar{n}_{j}$ and $\delta
_{j}=\delta \textrm{ mod }\bar{n}_{j}.$ Then the map%
\begin{equation}
\Gamma :SL(2,\bar{N})\rightarrow SL(2,\bar{n}_{1})\times \cdots \times SL(2,\bar{n}_{r}),
\end{equation}
defined by 
\begin{equation}
\Gamma (F)=\left( F_{1},\ldots ,F_{r}\right) ,
\end{equation}
is an isomorphism. The proof follows a similar argument to the above and
implies that 
\begin{eqnarray}
C(N) &\simeq &H(n_{1})\times \cdots \times H(n_{r})\times SL(2,\bar{n}_{1})\times \cdots \times SL(2,\bar{n}_{r}) \nonumber \\
&\simeq &H(n_{1})\times SL(2,\bar{n}_{1})\times \cdots \times H(n_{r})\times SL(2,\bar{n}_{r}) \\
&\simeq &C(n_{1})\times \cdots \times C(n_{r}). \nonumber
\end{eqnarray}

These two observations mean that every displacement operator can be written
as a tensor product of displacement operators in smaller Hilbert spaces
because of the following fact from finite group theory (see for example 
Theorem 19.18 of Ref. \cite{james+01}). Let $%
G$ and $J$ be groups, then \emph{every} irreducible representation of the
group $G\times J$ is a tensor product of an irreducible representation of $G$
with an irreducible representation of $J.$ The Weyl-Heisenberg and Clifford
groups are direct products of the groups defined over the prime
factorization and therefore all irreducible representations can be written
as a tensor product of irreducible representations of the smaller groups.

Care is required when writing down the isomorphisms in terms of the standard
unitary representation, $X$, $Z$ and $\tau ,$ defined in eq. (3). If we
define the map 
\begin{equation}
\eta :x^{a}z^{b}t^{c}\rightarrow \left( \tau _{1}^{c}\ldots \tau
_{r}^{c}\right) \left( X_{1}^{a}Z_{1}^{b}\otimes \cdots \otimes
X_{r}^{a}Z_{r}^{b}\right) ,  \label{map 1}
\end{equation}%
we have the problem that in general, 
\begin{equation}
\tau _{1}^{c}\ldots \tau _{r}^{c}\neq \tau ^{c}, 
\end{equation}
meaning that the right hand side of eq. (\ref{map 1}) cannot be the image of $%
\theta $ under any unitary induced mapping. To fix this problem, we redefine 
$\eta $ to be%
\begin{equation}
\eta ^{\prime }:x^{a}z^{b}t^{c}\rightarrow \left( \tau _{1}^{\kappa
_{1}c}\ldots \tau _{r}^{\kappa _{r}c}\right) \left(
X_{1}^{a}Z_{1}^{\kappa_{1}b}\otimes \cdots \otimes X_{r}^{a}Z_{r}^{\kappa_{r} b}\right) , 
\end{equation}
where $\kappa _{j}$ is the multiplicative inverse of $N/n_{j} \textrm{ mod }\bar{n%
}_{j}.$ The map $\eta ^{\prime }$ then satisfies all of the required
properties to be an isomorphism.

To construct the isomorphism for the standard unitary representation of the
Clifford group, we take the image of the symplectic matrices, $F_{j}$ to be $%
U_{F_{j}^{\prime }},$ where 
\begin{equation}
F_{j}^{\prime }=\left( 
\begin{array}{cc}
\alpha _{j} & \kappa^{-1} _{j}\beta _{j} \\ 
\kappa _{j}\gamma _{j} & \delta _{j}%
\end{array}%
\right) , 
\end{equation}
rather than $U_{F_{j}}.$ Whilst we did not need the explicit form of the two
isomorphisms in this paper, we hope that it will prove a useful tool
elsewhere.

\newpage

{\bf Appendix C: Fiducial vectors for $N=16$}

\vspace{5mm} 

\printfidsym{16a}
\printfidsym{16b}

\end{document}